\documentclass[twocolumn]{aastex63}
\begin{document}

\title{2~mm GISMO Observations of the Galactic Center. 
I. Dust Emission\footnote{Based on observations carried out with the IRAM 30~m Telescope. 
IRAM is supported by INSU/CNRS (France), MPG (Germany) and IGN (Spain).}}

\shortauthors{Arendt et al.}
\shorttitle{2mm Emission from Galactic Center Dust}

\author[0000-0001-8403-8548]{Richard G. Arendt} 
\affiliation{Code 665, NASA/GSFC, 8800 Greenbelt Road, Greenbelt, MD 20771, USA}
\affiliation{CRESST2/UMBC}

\author[0000-0002-8437-0433]{Johannes Staguhn}
\affiliation{Code 665, NASA/GSFC, 8800 Greenbelt Road, Greenbelt, MD 20771, USA}
\affiliation{Johns Hopkins University}

\author[0000-0001-8033-1181]{Eli Dwek} 
\affiliation{Code 665, NASA/GSFC, 8800 Greenbelt Road, Greenbelt, MD 20771, USA}

\author[0000-0002-6753-2066]{Mark R. Morris} 
\affiliation{Department of Physics and Astronomy, University of California Los Angeles, Los Angeles, CA 90095, USA}

\author{Farhad Yusef-Zadeh} 
\affiliation{CIERA and the Department of Physics \& Astronomy, Northwestern University, 2145 Sheridan Road, Evanston, IL 60208, USA}

\author[0000-0002-9884-4206]{Dominic J. Benford} 
\affiliation{Astrophysics Division, NASA Headquarters, 300 E St. SW, Washington, DC 20546, USA}

\author[0000-0001-8991-9088]{Attila Kov\'acs} 
\affiliation{Smithsonian Astrophysical Observatory Submillimeter Array (SMA), MS-78, 60 Garden St, Cambridge, MA 02138, USA}

\author{Junellie Gonzalez-Quiles} 
\affiliation{Code 667, NASA/GSFC, 8800 Greenbelt Road, Greenbelt, MD 20771, USA}
\affiliation{CRESST2/SURA}

\email{Richard.G.Arendt@nasa.gov,
Johannes.G.Staguhn@nasa.gov}
\email{Eli.Dwek@nasa.gov,
morris@astro.ucla.edu}
\email{zadeh@northwestern.edu,
Dominic.J.Benford@nasa.gov}
\email{attila.kovacs@cfa.harvard.edu,
Junellie.Gonzalez-Quiles@nasa.gov}

\begin{abstract}
The Central Molecular Zone (CMZ), covering the 
inner $\sim 1\arcdeg$ of the Galactic plane has been mapped at 2~mm using
the GISMO bolometric camera on the 30~m IRAM telescope. 
The $21''$ resolution maps show abundant emission from cold molecular clouds,
from star forming regions, and from one of the Galactic center nonthermal 
filaments. In this work we use the Herschel Hi-GAL data to model the dust
emission across the Galactic center. 
We find that a single-temperature fit can describe
the 160 -- 500 \micron\ emission for most lines of sight, if the long-wavelength 
dust emissivity scales as $\lambda^{-\beta}$ with $\beta \approx 2.25$. This
dust model is extrapolated to predict the 2~mm dust emission. Subtraction 
of the model from the GISMO data provides a clearer look at the 2~mm emission 
of star-forming regions and the brightest nonthermal filament.
\end{abstract}


\section{Introduction}

\begin{figure*}[t] 
   \centering
   \includegraphics[width=6.5in]{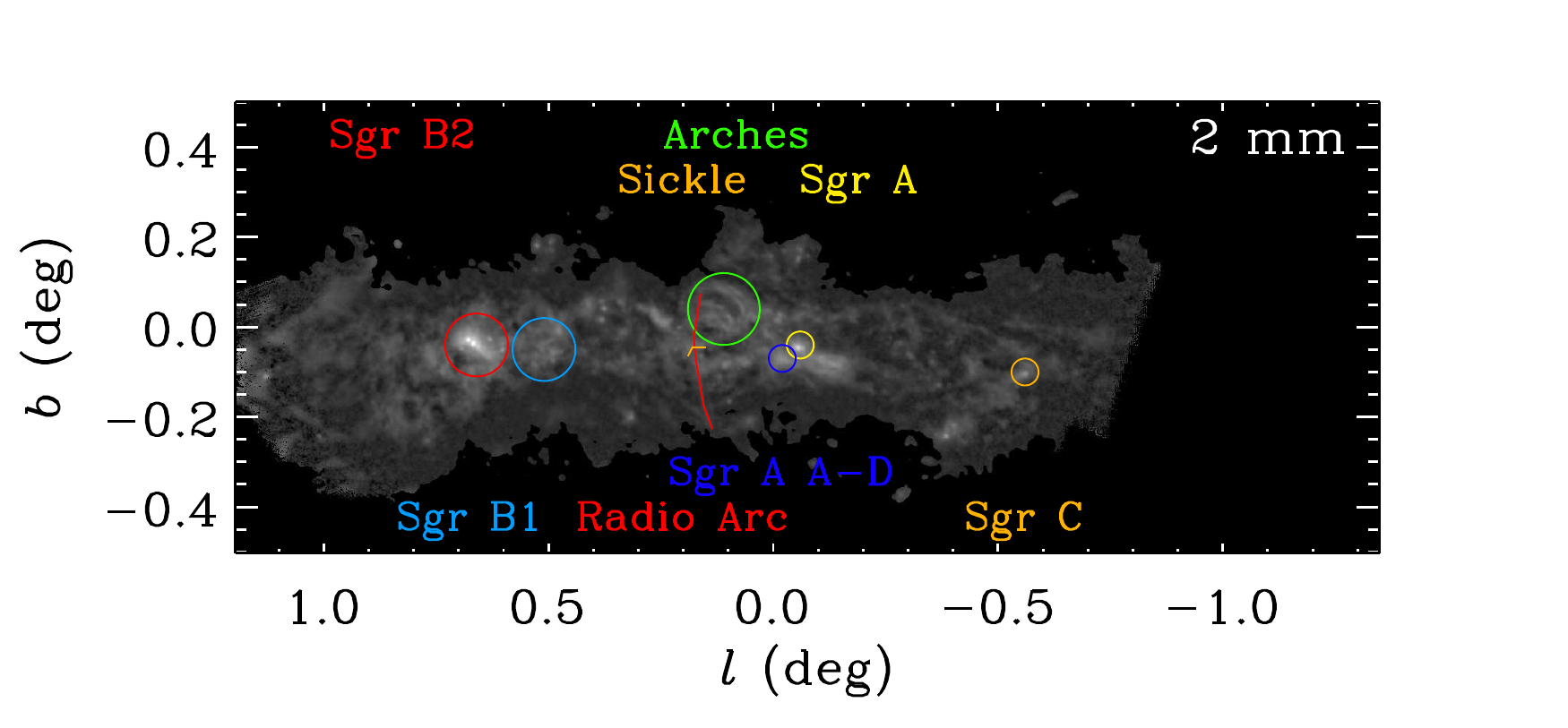} \\
   \caption{Features in the central molecular zone. Originally defined from 
   longer wavelength radio observations, these features are notable 2~mm  
   sources as well. See Figure \ref{fig:gismo} for an unadorned version 
   of the 2~mm image.
   \label{fig:guide}}
\end{figure*}

The Galactic Center presents an interesting challenge to observe and
understand because of the huge variety of sources and structures that 
are superimposed along the line of sight.
The line of sight extinction prevents any useful observations at UV or 
optical wavelengths. The development of infrared, radio, and X-ray 
observations has provided a clearer view of the Galactic center region. 
Infrared and radio observations have revealed a central molecular zone (CMZ)
within Galactic longitude $|l|\lesssim1\arcdeg$ or a radius 
of $\sim140$~pc of the Galactic center \citep{Morris:1996}.
The CMZ contains the largest, densest collection of giant molecular 
clouds in the Galaxy. Originating from these clouds are numerous 
star forming regions and young stellar objects 
with various sizes and ages \citep[e.g][]{Yusef-Zadeh:2009}. These in turn have 
generated large stellar clusters, including the nuclear cluster that 
swarms around the Galaxy's central supermassive black hole, and the 
more outlying Arches and Quintuplet clusters 
\citep[e.g.][]{Figer:1999a,Figer:2002,Genzel:2003}.
The clusters and individual stars 
interact with their surrounding medium to produce ionized structures and 
bubbles \citep{Simpson:1997,Cotera:2005,Simpson:2007}. 
Non-thermal radio structures are seen in the form of supernova 
remnants and an assortment of individual and bundles of filaments.
High-energy (X-ray) observations \citep[e.g.][]{Wang:2002} 
reveal emission from sources such as 
accreting compact objects (white dwarfs, neutron stars, 
or stellar-mass black holes) in binary systems, very hot gas, 
and colliding stellar winds \citep{Wang:2006,Yusef-Zadeh:2015}. 
Figure \ref{fig:guide} provides a guide to the locations of 
some of the relevant structures in the CMZ.

New observations with new instruments invariably provide insight to 
one or more of these components of the Galactic Center region. 
We have carried out a 2~mm continuum survey of the CMZ region 
using the IRAM 30~m telescope \citep{Baars:1987} paired with the 
Goddard-IRAM Superconducting 2-Millimeter Observer (GISMO) instrument
\citep{Staguhn:2006,Staguhn:2008}.
Observations at 2~mm are in an interesting transition zone between 
mid- and far-IR wavelengths on the one hand
(10-500 \micron), which are generally dominated by emission from warm and cold 
interstellar dust, and radio wavelengths on the other hand ($>1$ cm),
where the dominant 
emission mechanisms are free-free emission from ionized gas and synchrotron
emission from relativistic electrons.
The GISMO survey offers coverage of a wide angular extent (the entire CMZ) 
with excellent sensitivity and the ability to trace large angular scale 
emission that higher angular resolution interferometric observations miss.
Previous similar surveys and observations are: the BOLOCAM Galactic 
Plane Survey \citep{Bally:2010,Ginsburg:2013} which covers the entire CMZ
at 1.1~mm and is thus dominated by cold dust emission;  
3~mm interferometric observations of the inner portion of the CMZ from 
Sgr A to the Radio Arc by \cite{Pound:2018}, which are sensitive to 
thermal and non-thermal radio emission and very insensitive to 
thermal dust emission; 2~mm observations of the Radio Arc region 
by \cite{Reich:2000}; and relatively low angular resolution observations 
at 2 and 3~mm (and other wavelengths) by cosmic microwave background surveys 
\citep{Culverhouse:2010,Planck-Collaboration:2015}.

In this paper we present the full 2~mm GISMO image of the Galactic center (Section 2).
Here we concentrate on understanding the thermal emission from cold dust in 
molecular clouds which makes up the bulk of the emission seen at 2~mm. 
In section 3 we describe the modeling of the {\it Herschel} far-IR observations 
of the CMZ region for the purpose of extrapolating this emission to 2~mm. 
The comparison of the extrapolated dust emission to the observed 2~mm 
emission provides a check on the inferred dust properties, temperature and 
emissivity, and a means to separate 2~mm emission that arises from sources 
other than cold dust. The comparison between 2~mm free-free emission and 
thermal emission from dust allows us to make rough estimates of the gas
density in regions where these components coexist (Section 4). 
Investigation of 2~mm emission from the brightest of the nonthermal radio 
filaments in the Galactic center is presented in a separate paper \citep{Staguhn:2019}.

\section{Data and Resolution} \label{sec:data}

\subsection{GISMO}\label{ssec:gismo}
The 2~mm GISMO bolometric data were collected during April and November 2012.
A raster scanning pattern was used for 52 separate scans that are combined here.
The total observing time was approximately 8 hours and results in 
a median integration time of $\sim1.1$ s~pixel$^{-1}$.  
CRUSH (version 2.22-b1) was used for the data reduction \citep{Kovacs:2008},
employing the ``faint'' and ``extended'' processing flags and 40 ``rounds'' of
iterations to help recover extended emission.
The final image has a median noise level of 4.3 mJy beam$^{-1}$.
The data reduction yields an image spanning roughly
$1.2\arcdeg>l>-0.8\arcdeg$ and $|b| < 0.3$, 
with an effective Gaussian beam of $21''$ (FWHM).
Although these data are collected from a
single-dish telescope (not an interferometer), the data reduction 
necessitated some loss of large-scale structure due to the time-domain 
filtering required to remove atmospheric variations. There is also no 
measure of the absolute zero point of the intensity. The GISMO
data have been converted from the default units of Jy beam$^{-1}$ 
to MJy sr$^{-1}$ to facilitate comparison with the IR maps.

\subsection{Herschel}

The {\it Herschel} data used are data release 1 (DR1) data products, 
collected and processed as part of the Herschel infrared Galactic Plane Survey 
(Hi-GAL) survey \citep{Molinari:2016}. 
The data are broad-band images from both the PACS
\citep{Poglitsch:2010} and SPIRE \citep{Griffin:2010} instruments.
The nominal wavelengths and angular resolution of the data are 
listed in Table \ref{tab:data}. The data were obtained in seven 
separate fields from the archive\footnote{\url{http://vialactea.iaps.inaf.it/vialactea/eng/index.php}} 
and re-joined together to cover the entire region observed by GISMO.

\subsection{Convolution}
To avoid artifacts in the modeling, we used the procedures of 
\cite{Aniano:2011} to convolve all the images to a common resolution of 
$37''$ (FWHM) which is limited by the 500 \micron\ {\it Herschel} SPIRE data
(see Table \ref{tab:data}). 
For some later investigation of specific features in the 2~mm image, we 
convolved images to match GISMO's $21''$ resolution. In such cases one
must be cognizant of possible artifacts due to mismatched resolutions 
of the 350 and 500 $\mu$m data. 

\begin{figure*}[t] 
   \centering
   \includegraphics[width=6.5in]{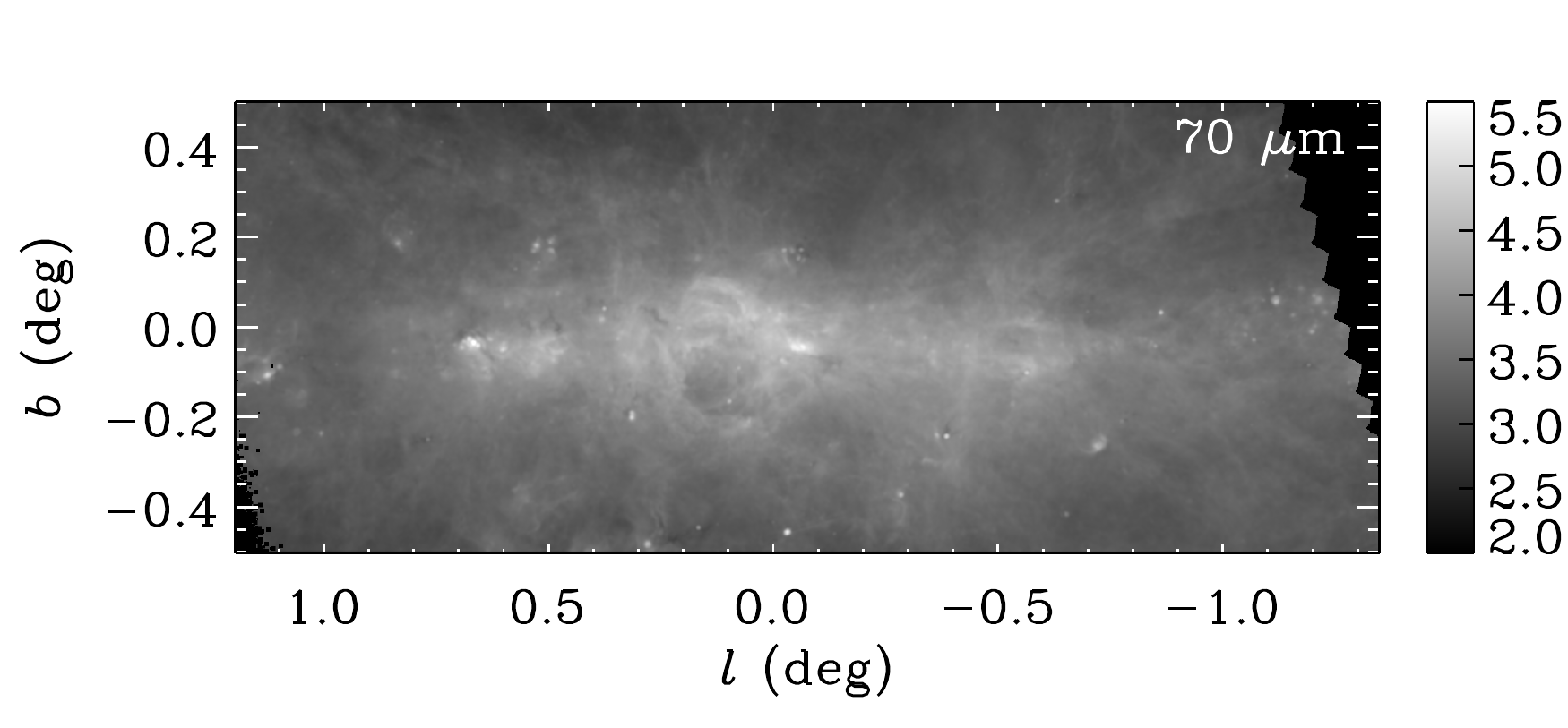} \\
   \includegraphics[width=6.5in]{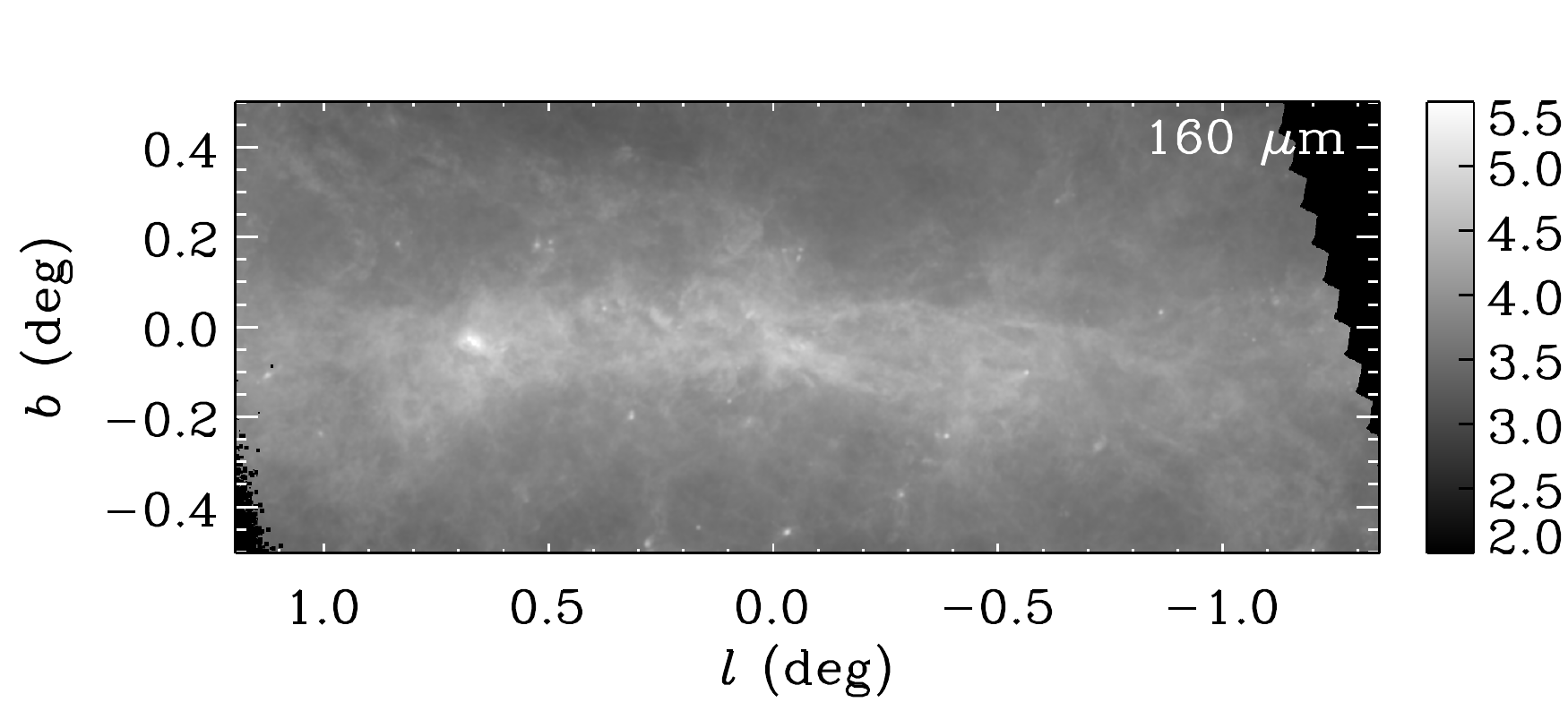}
   \caption{{\it Herschel} PACS 70 (top) and 160 $\mu$m (bottom) 
   images from the Hi-GAL survey 
   \citep{Molinari:2016}. These are displayed on a logarithmic scale spanning 
   $[100,3.5\times10^5]$ MJy sr$^{-1}$. These images are convolved to the 
   $21''$ resolution of the GISMO 2~mm data. Emission is dominated by cool 
   dust ($T_{\rm d} \approx 19$~K, see Figure \ref{fig:free_beta}), 
   but warmer dust components ($T_{\rm d}$ up to 30~K) in compact and extended 
   \ion{H}{2} regions are evident at 70~$\mu$m.
   \label{fig:pacs}}
\end{figure*}

\begin{figure*}[t] 
   \centering
   \includegraphics[width=6.5in]{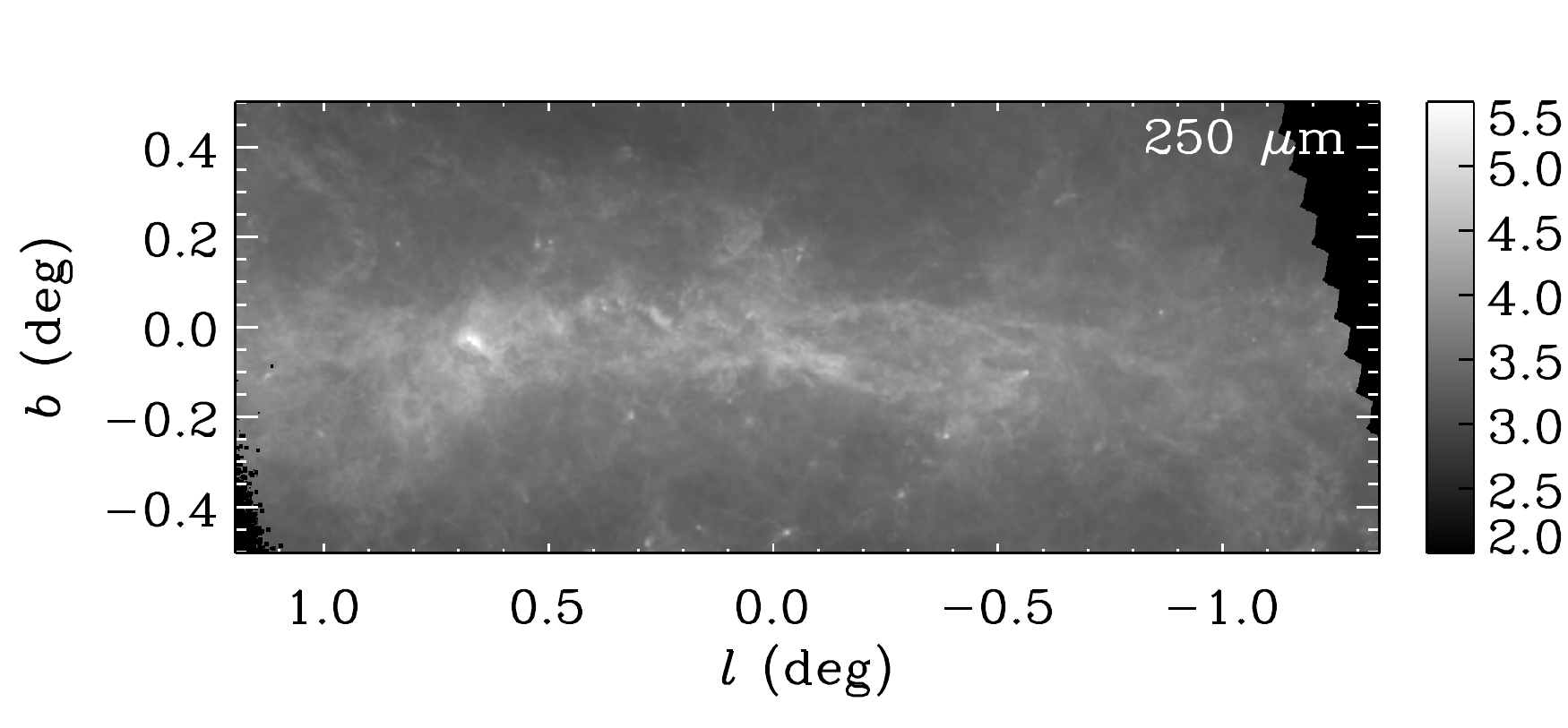} \\
   \includegraphics[width=6.5in]{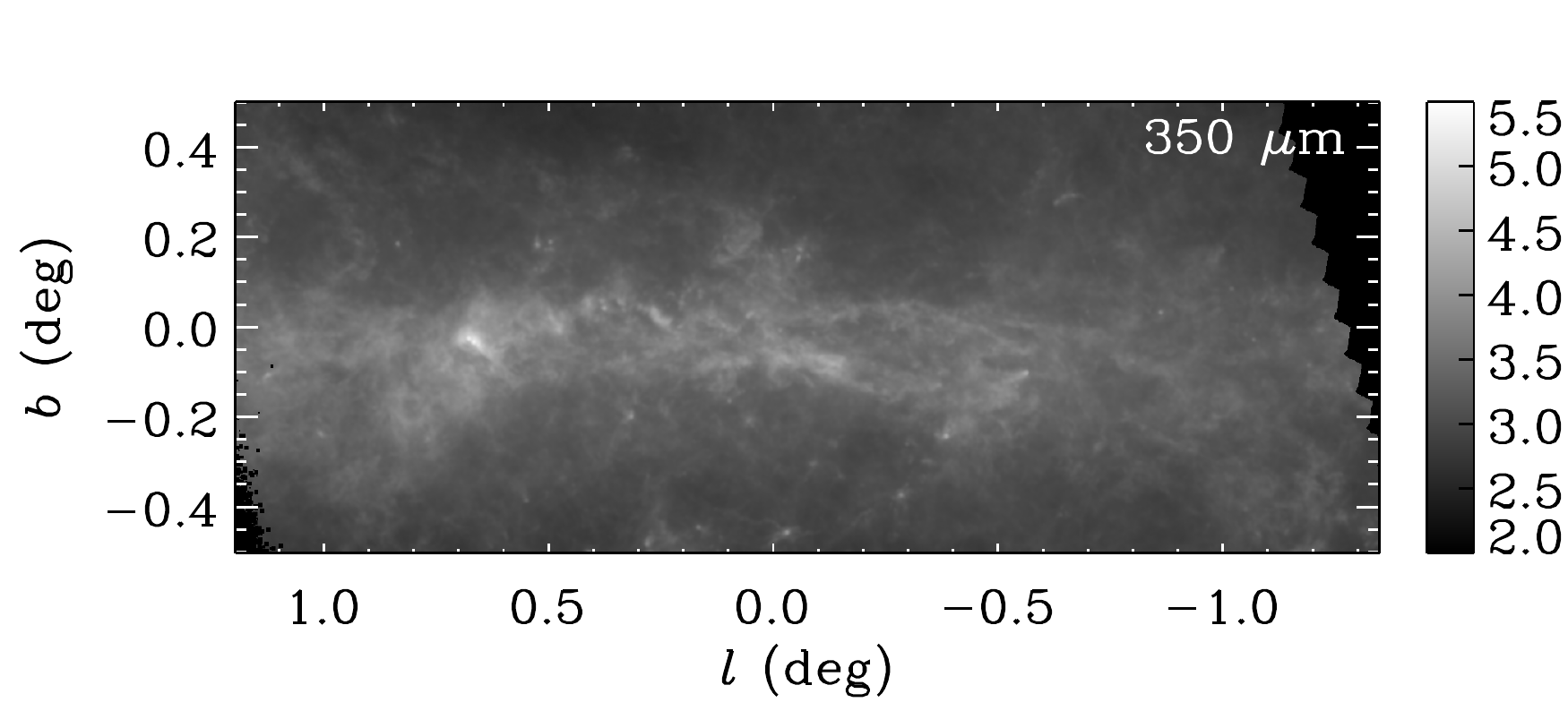} 
   \caption{{\it Herschel} SPIRE 250 (top) and 350~$\mu$m (bottom)
   images from the Hi-GAL survey 
   \citep{Molinari:2016}. These are displayed on a logarithmic scale spanning 
   $[100,3.5\times10^5]$ MJy sr$^{-1}$. The 250 $\mu$m image is convolved to the 
   $21''$ resolution of the GISMO 2~mm data. Emission is dominated by cool 
   dust ($T_{\rm d} \approx 19$~K, see Figure \ref{fig:free_beta}).
   \label{fig:spire}}
\end{figure*}

\begin{figure*}[t] 
   \centering
   \includegraphics[width=6.5in]{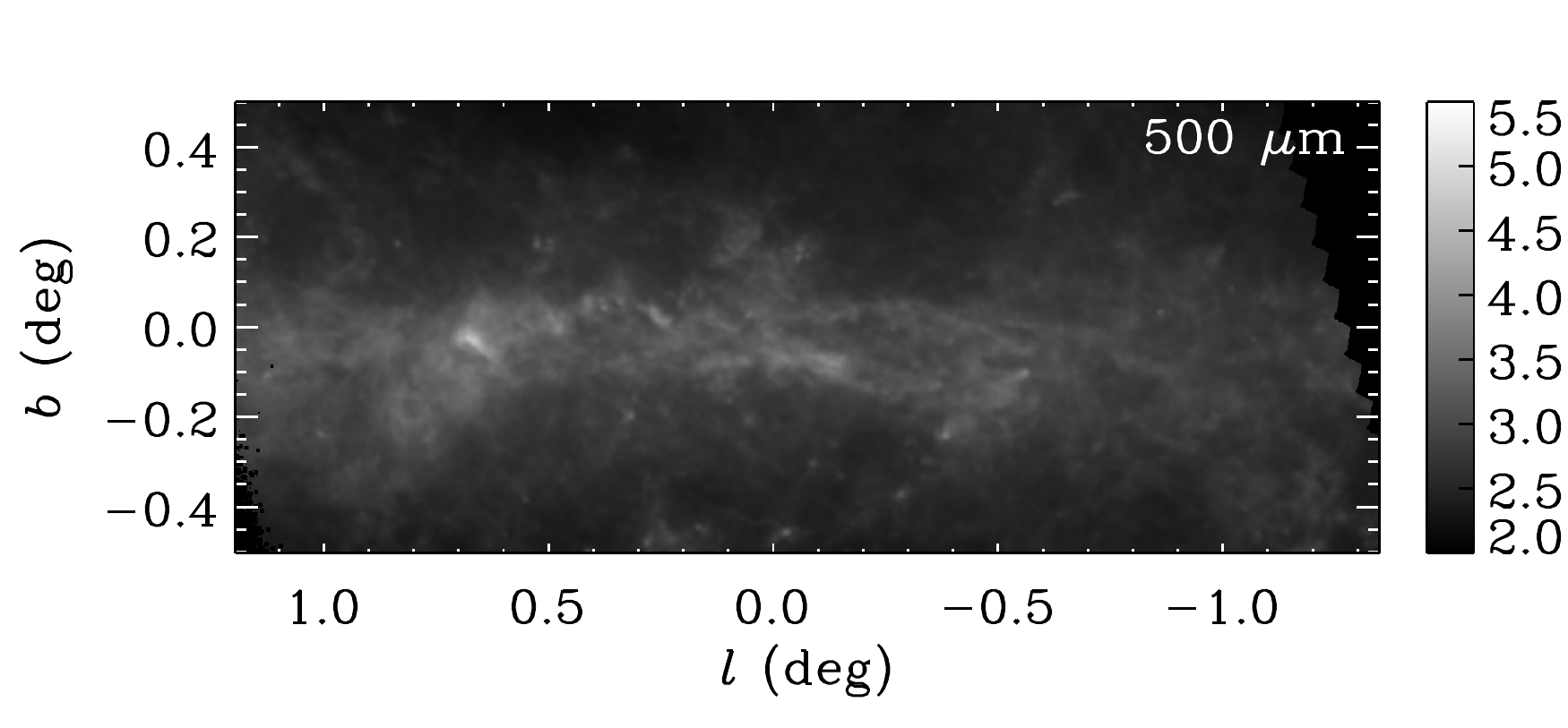} \\
   \includegraphics[width=6.5in]{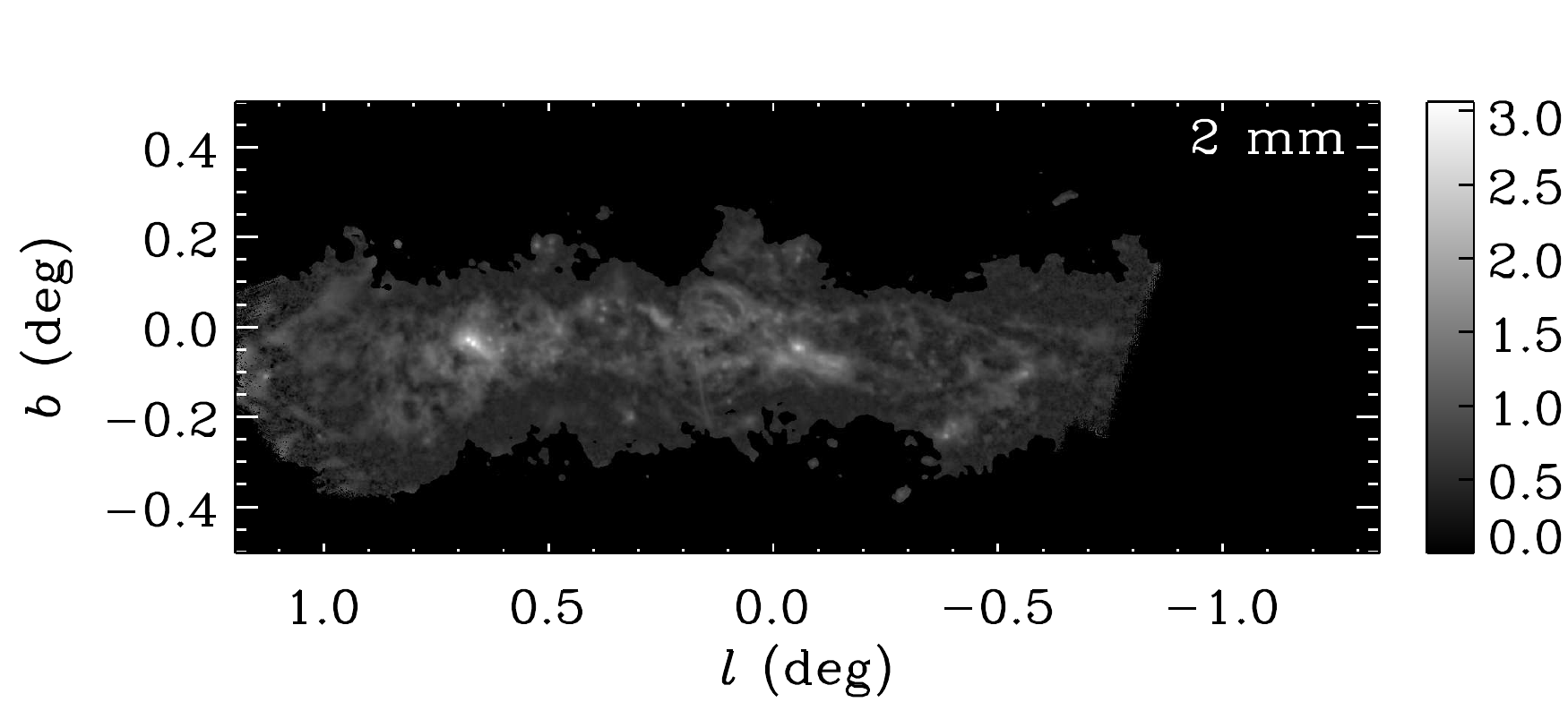} \\
   \includegraphics[width=6.5in]{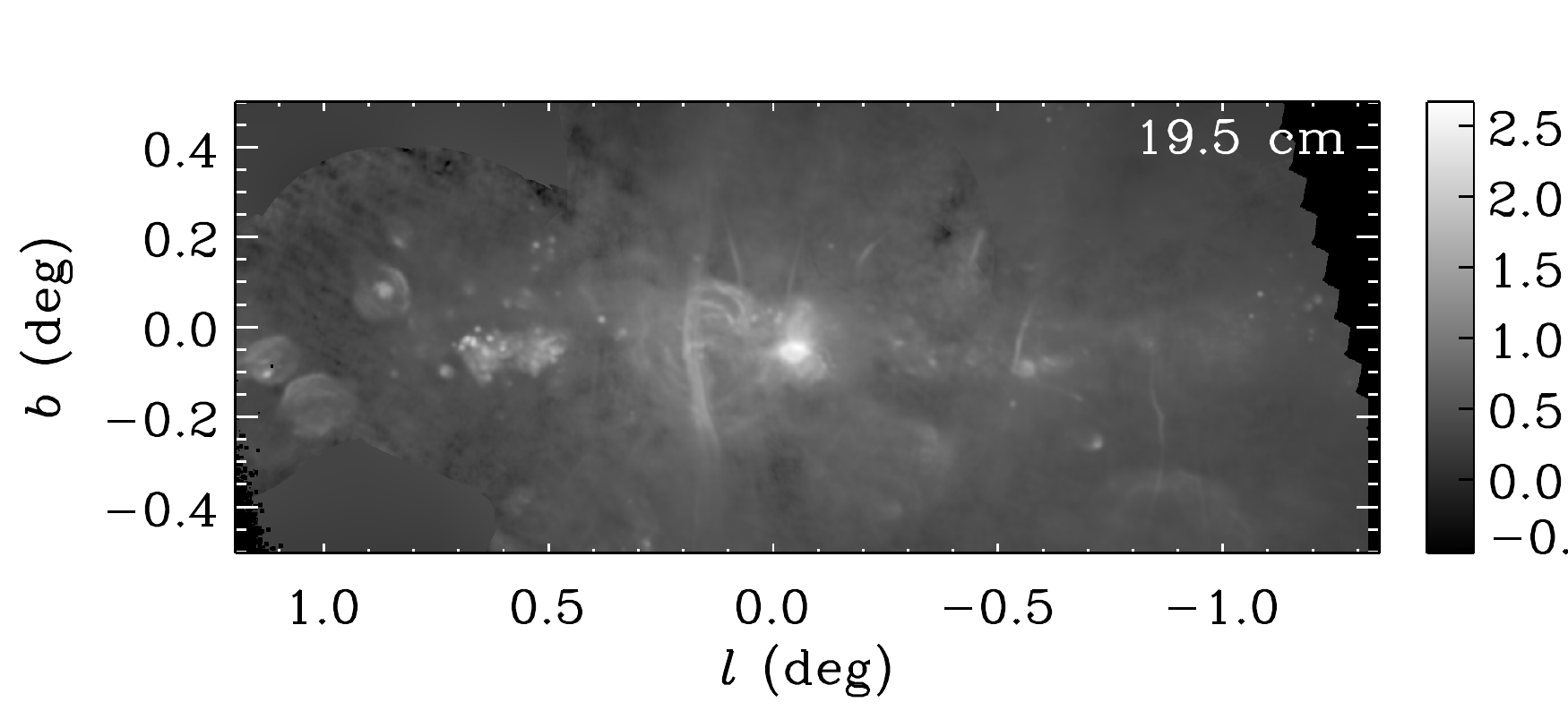}
   \caption{ (top to bottom) SPIRE 500~$\mu$m, GISMO 2~mm and VLA 19.5 cm images 
   of the Galactic center region. These are displayed on a logarithmic 
   scale spanning $[100,3.5\times10^5]$, $[1,1150]$, and 
   $[0.3,470]$ MJy sr$^{-1}$ respectively (an offset of +3 MJy sr$^{-1}$
   is applied to the 2~mm image 
   to avoid truncating the negative noise variations of the background).
   The 19.5~cm image exhibits both thermal (free-free) and 
   non-thermal (synchrotron) sources. The 2~mm image shows features 
   in common with both longer and shorter wavelength images.
   \label{fig:gismo}}
\end{figure*}

The Hi-GAL and GISMO images at $21''$ resolution are shown in 
Figures \ref{fig:pacs}-\ref{fig:gismo}. Figure \ref{fig:gismo} also shows the 19.5 cm  radio 
continuum image of the region from \citet{Yusef-Zadeh:2009} which reveals free-free and synchrotron emission.

\begin{deluxetable}{lccl}
\tablewidth{0pt}
\tablecaption{Data Sets Used\label{tab:data}}
\tablehead{
\colhead{Instrument} & 
\colhead{Wavelength (\micron)} & 
\colhead{FWHM ($''$)} 
}
\startdata
PACS & 70 & 5.2\\
PACS & 160 & 12\\
SPIRE & 250 & 18\\
SPIRE & 350 & 25\\
SPIRE & 500 & 37\\
GISMO & 2~mm & 21\\
\enddata
\end{deluxetable}

\section{Dust Emission Models} \label{sec:analysis}

Inspection of 
Figure \ref{fig:gismo} 
shows that much of the 2~mm emission is
correlated with far-IR (500 $\mu$m) emission. 
However, there are additional features
in the 2~mm image that instead correlate with the radio (19.5 cm) emission. 
In order to characterize each of these emission components, we begin by 
constructing a model of the dust emission based on the far-IR {\it Herschel} 
data. This model can be used to predict and subtract the dust emission 
at 2~mm.

\subsection{Far-IR emission} \label{sec:far-ir}

The observed surface brightness, $I_{\nu}$, of any pixel is given by 
\begin{equation}
I_{\nu} = \int_0^{\tau_\nu} B_{\nu}(T'_{\rm d}) e^{-\tau'_{\nu}} d\tau'_{\nu}
\end{equation}
where $B_{\nu}(T'_{\rm d})$ is the Planck function,
$T'_{\rm d}$ and $\tau'_{\nu}$ are the temperature and 
optical depth along the line of sight, 
and $\tau_\nu$ is the total optical depth on the line of sight.
Initial examination of the {\it Herschel} data indicated that 
the 160-500 \micron\ data could generally be accurately fit if 
$T'_{\rm d}$ is taken to be constant ($T_{\rm d}$) along the line of sight,
while varying for each different line of sight. 
Therefore we can model the emission as
\begin{eqnarray}
I_{\nu} & = & B_{\nu}(T_{\rm d})\ \int_0^{\tau_\nu} e^{-\tau'_{\nu}} d\tau'_{\nu}\\
 & = & B_{\nu}(T_{\rm d})\ (1-e^{-\tau_{\nu}})\\
 & \approx & B_{\nu}(T_{\rm d})\ \tau_{\nu}\\
 & = & B_{\nu}(T_{\rm d})\ {\cal M}_d\ \kappa_{\nu}.
\end{eqnarray}
With Equation (4) we assume $\tau_\nu \ll 1$, 
and then express the total optical depth in terms of the
mass column density of the dust, ${\cal M}_d$, and the mass 
absorption coefficient of the dust, $\kappa_{\nu}$. At long wavelengths, 
it is common to characterize the dust emissivity with a power law, such that 
$\kappa_{\nu}(\lambda) = \kappa_0\ \lambda^{-\beta}$.

Therefore the model applied to fit the observed emission at 
$\lambda_i \in [160,\ 250,\ 350,\ 500]$ \micron\ is:
\begin{equation}
I_{\nu}(\lambda_i) = A\ B_{\nu}(T_{\rm d},\lambda_i)\ \lambda_i^{-\beta}
\end{equation}
where the parameters to be determined by $\chi^2$ minimization are:
the dust temperature, $T_{\rm d}$, the spectral index of the dust emissivity, $\beta$, 
and the normalization coefficient, $A$, which will be proportional to the product,
${\cal M}_d\ \kappa_{0}$. The fitting procedure 
integrates the model over the appropriate filter bandwidths
before comparison to the observed intensities.

\begin{figure*}[p] 
   \centering
   \includegraphics[width=6.5in]{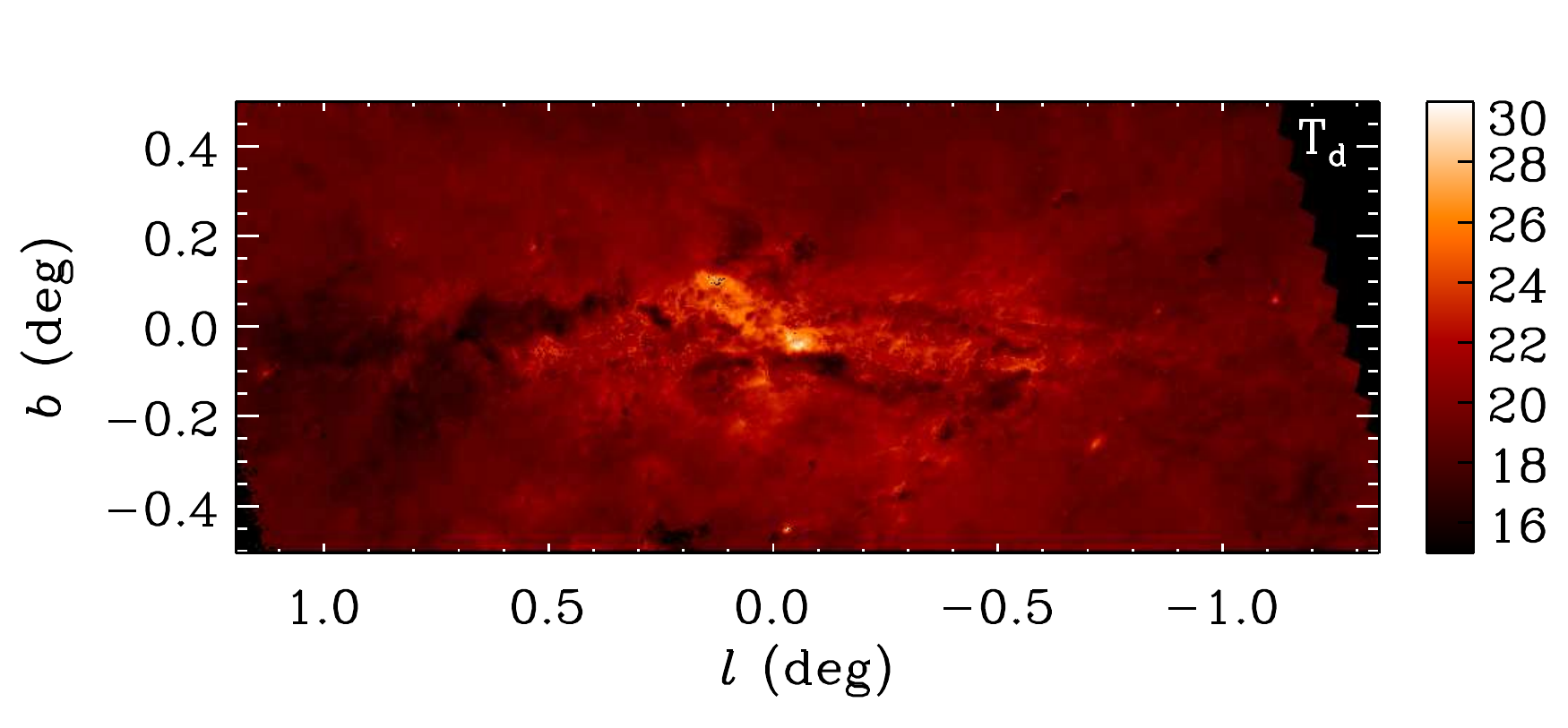}\\
   \includegraphics[width=6.5in]{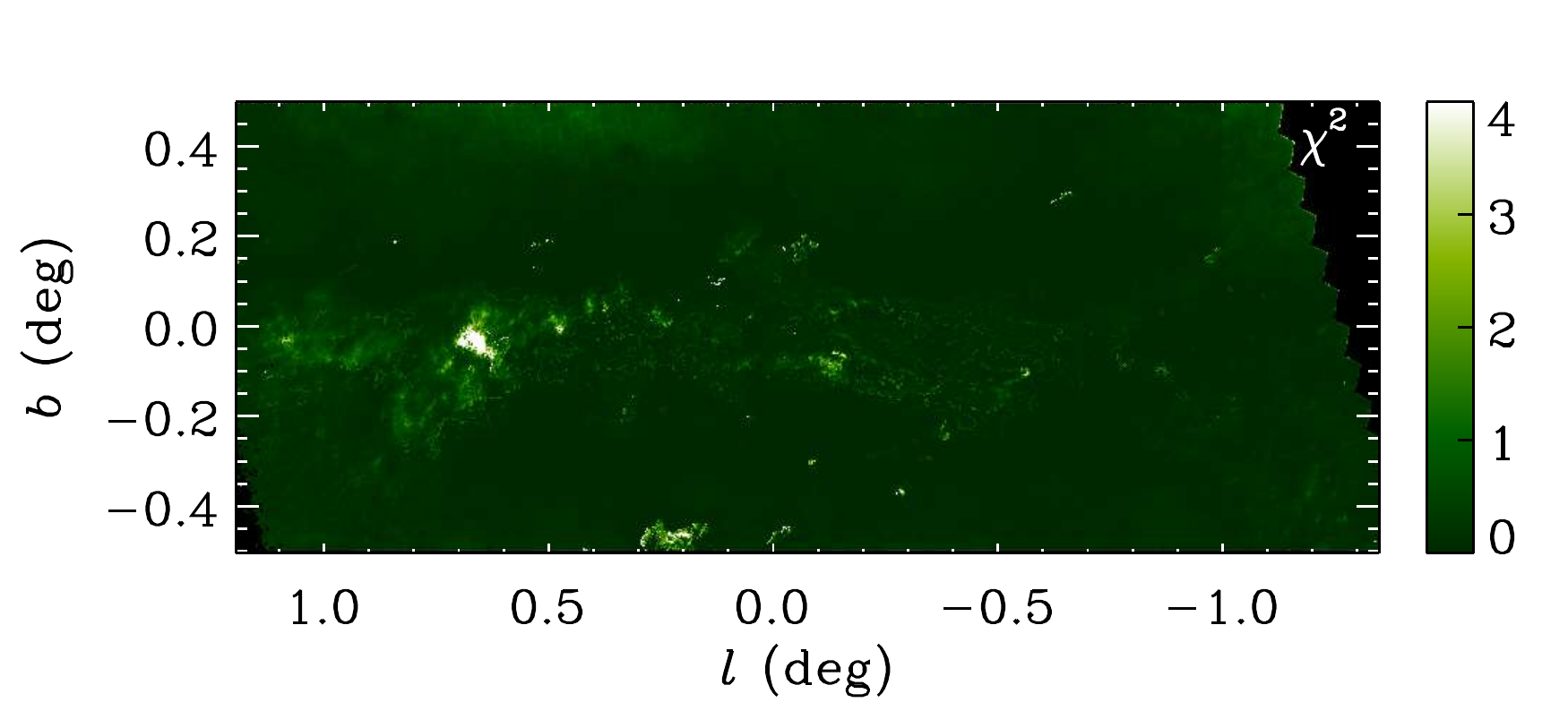}\\
   \includegraphics[width=6.5in]{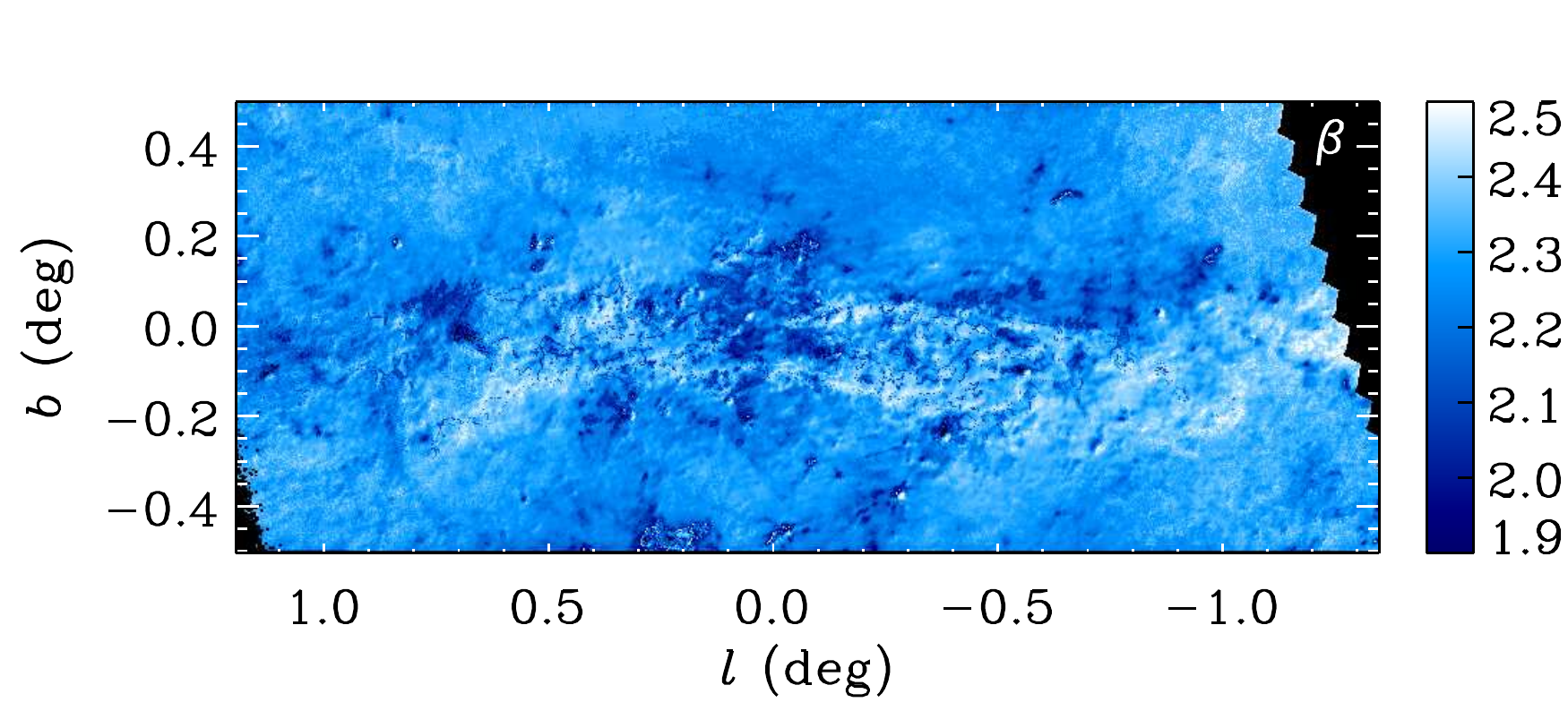} 
   \caption{Modeling the 160--500 $\mu$m images using both dust temperature 
   $T_{\rm d}$ (top) and $\beta$ (bottom) as free parameters. The display ranges 
   are [15,35] for $T_{\rm d}$ and [1.9,2.5] for $\beta$. The middle panel 
   shows $\chi^2$ for the fit at each pixel on a range of [0,4] 
   (fixed for comparison with later figures). Input data were convolved to 
   $37''$ resolution. $T_{\rm d}$ shows cold molecular clouds and warmer regions
   where strong local heating occurs. Results for $\beta$ are more erratic and 
   show spurious noise-induced features at the pixel scale.
   \label{fig:fit12}}
\end{figure*}

\begin{figure}[t] 
   \centering
   \includegraphics[width=3.35in]{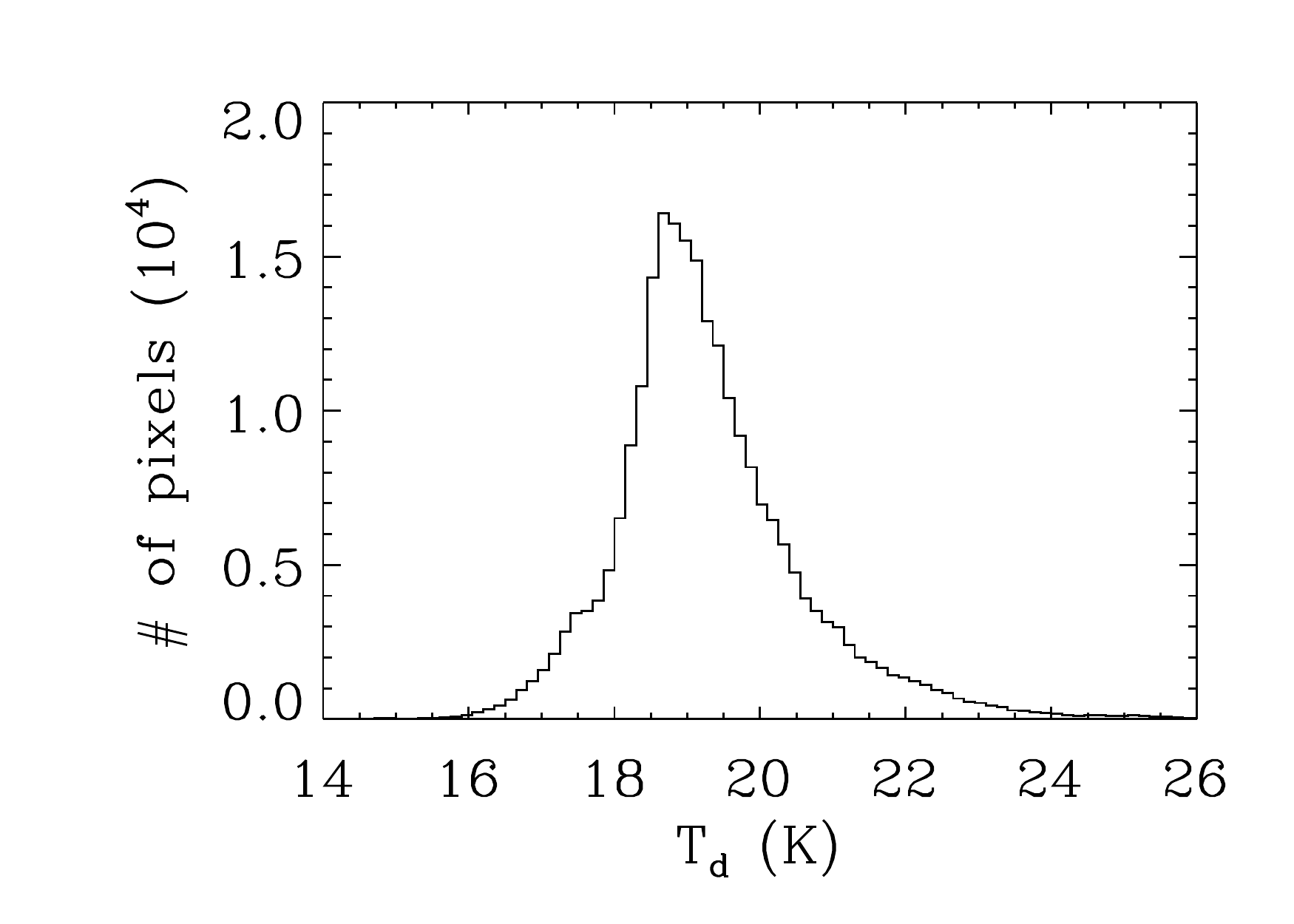}
   \includegraphics[width=3.35in]{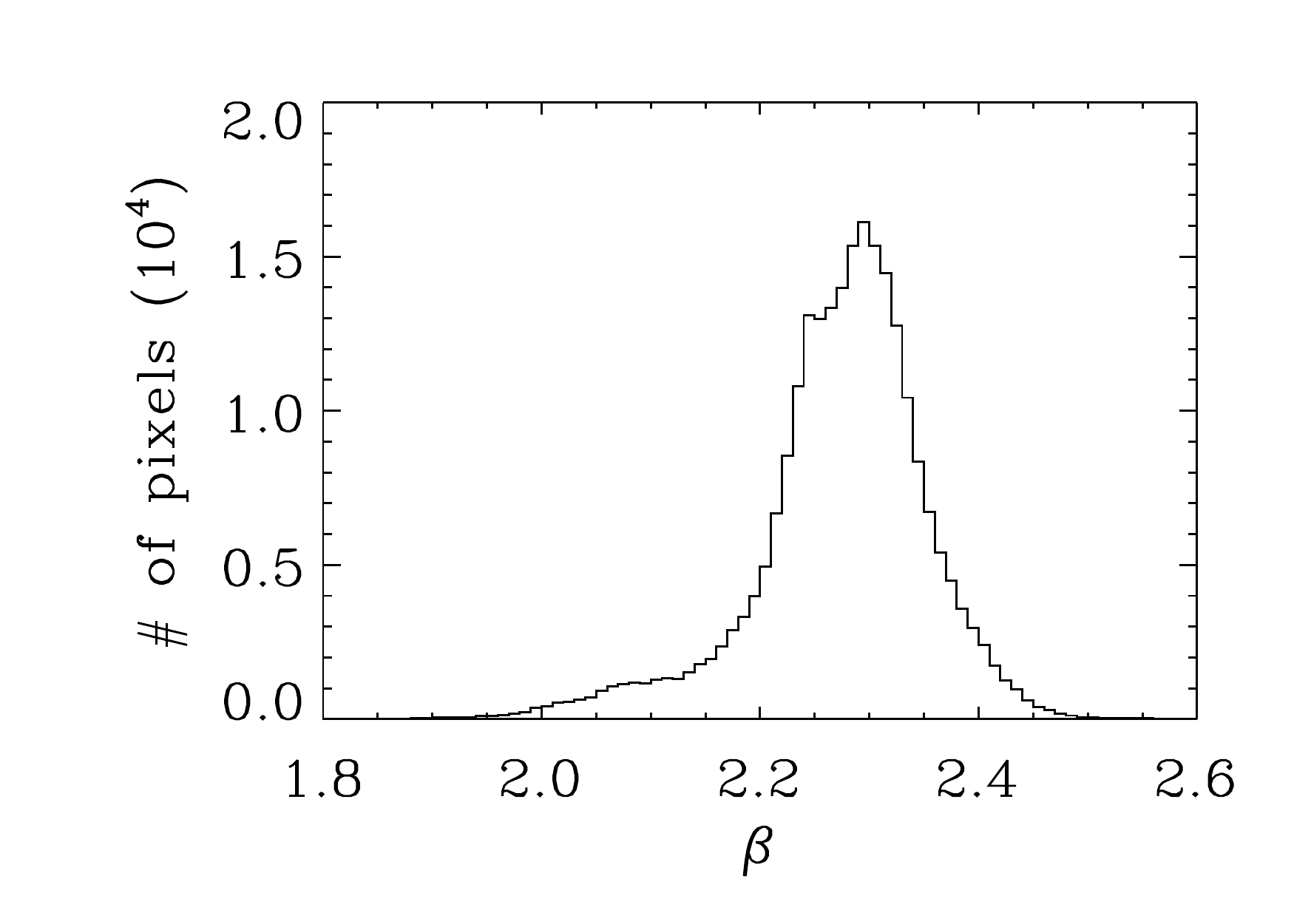}
   \caption{Histograms of the derived $T_{\rm d}$ (top) 
   and $\beta$ (bottom) for each pixel derived using the model in which 
   both are free parameters (See Figure \ref{fig:fit12}). 
   The mean and median values of $\beta$ are 2.27 and 2.28.
   Based on this distribution, we used a fixed $\beta = 2.25$ for 
   subsequent models.
   \label{fig:free_beta}}
\end{figure}

\begin{figure*}[t] 
   \centering
   \includegraphics[width=3.35in]{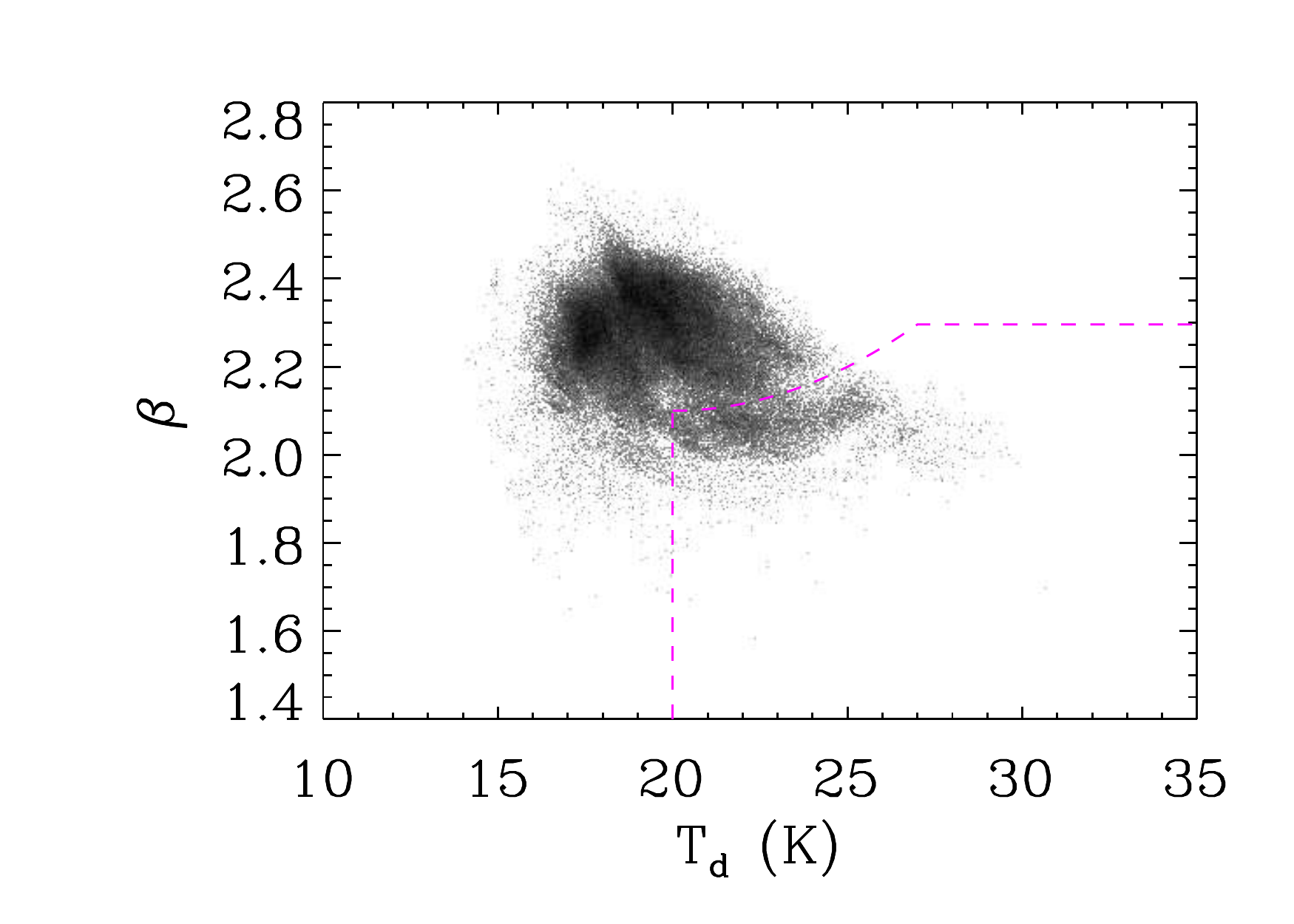}\\
   \includegraphics[width=6.5in]{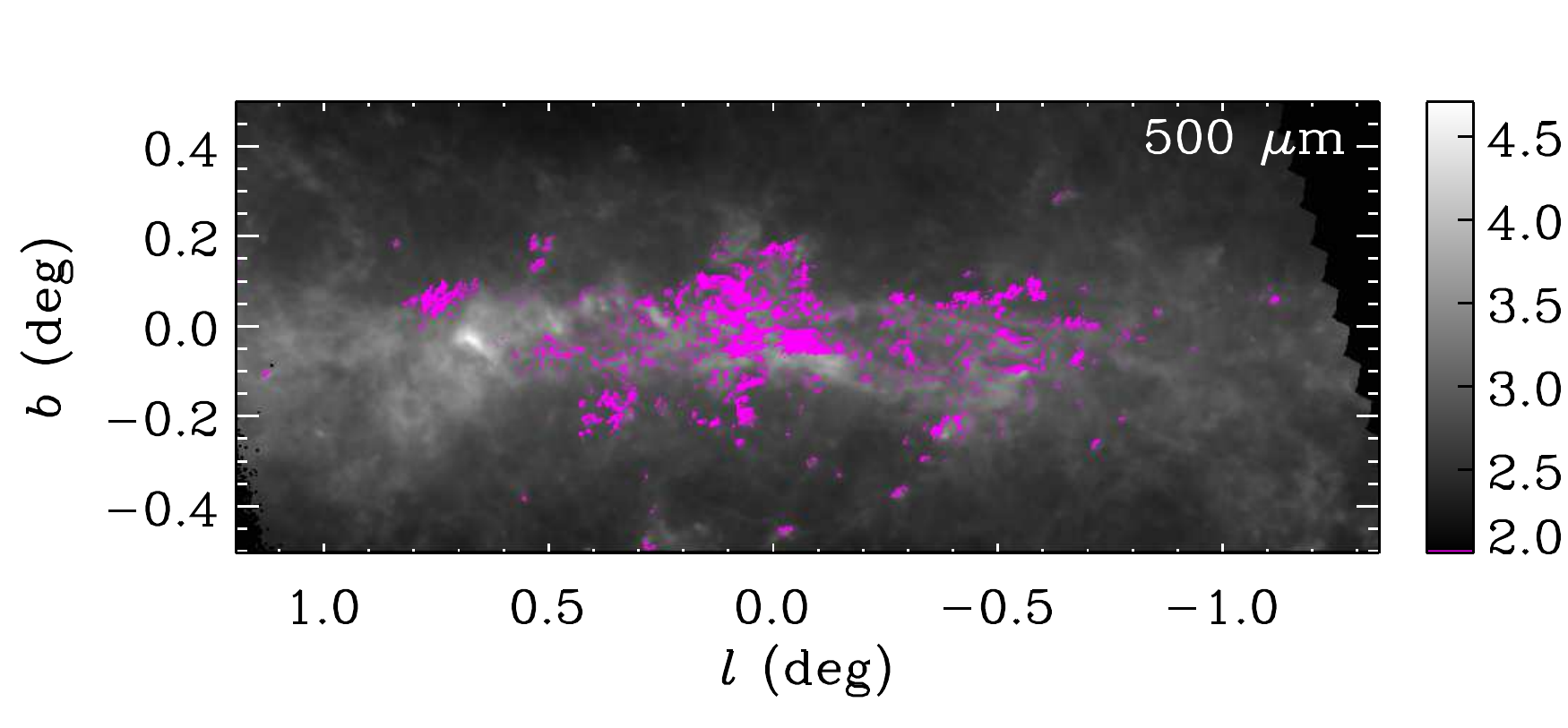}\\
   \includegraphics[width=6.5in]{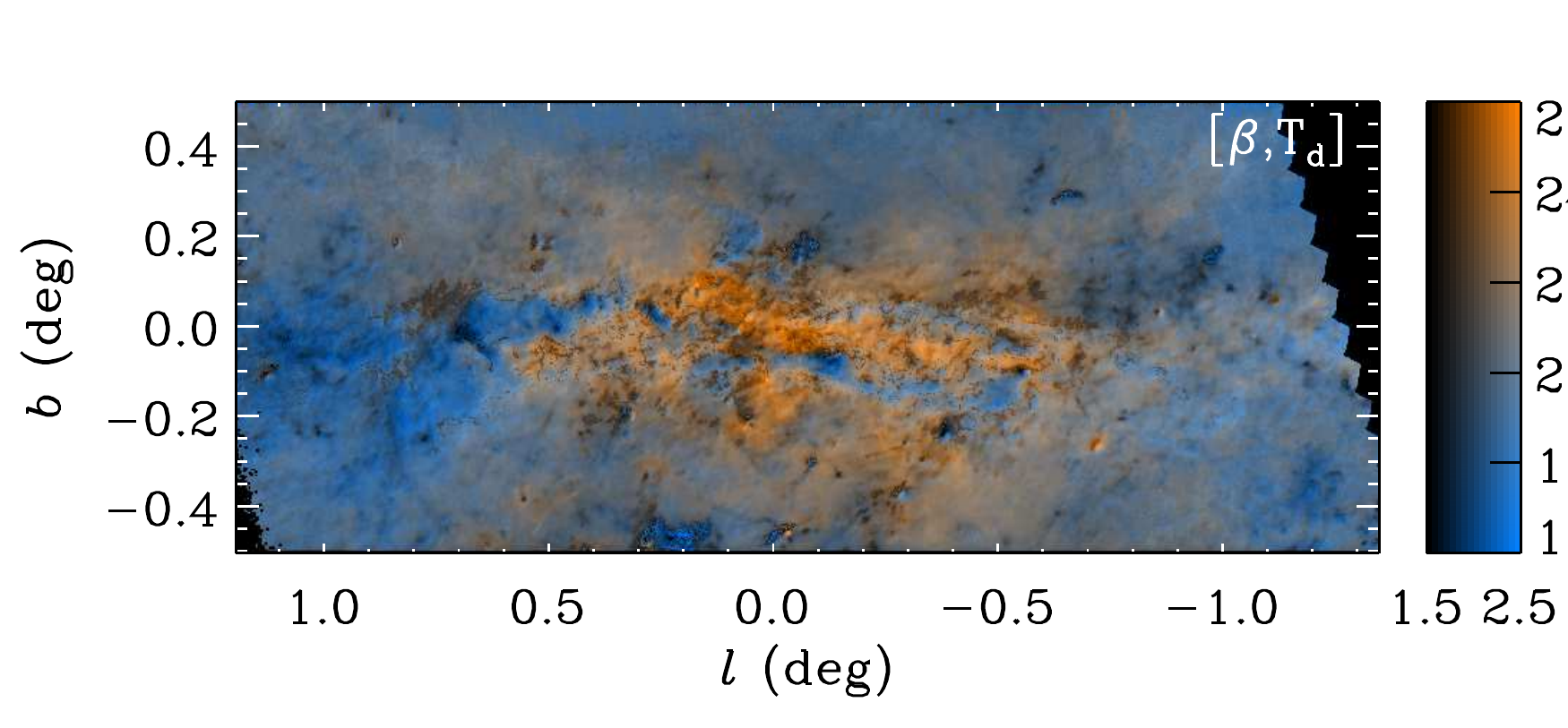}
   \caption{The top panel shows a 
   scatter plot between $T_{\rm d}$ and $\beta$ for all pixels shown 
   in Figure \ref{fig:fit12}. There is no tight correlation between these 
   parameters, although there is the loose trend that regions with high 
   temperatures are likely to have relatively low values of $\beta$.
   The magenta line separates a branch of points with high $T_{\rm d}$ and 
   low $\beta$ whose locations are marked in magenta on the 500~$\mu$m 
   image in the second panel.
   The bottom panel shows the simultaneous
   distribution of $T_{\rm d}$ and $\beta$, encoding $T_{\rm d}$ with 
   color from blue (cold) to orange (warm) and $\beta$ with brightness from
   low (dark) to high (light) (cf. Fig.~\ref{fig:fit12}).
   \label{fig:free_temp_beta}}
\end{figure*}

Figure \ref{fig:fit12} shows images of the dust temperature ($T_{\rm d}$) and the 
dust emissivity index ($\beta$) as derived from these fits. The image of $\chi^2$
of the fits is also shown. The image for the normalization ($A$) is very 
similar to that of $I_{\nu}(500\ \micron)$ and is omitted.
On large scales, the derived dust temperature is higher than average in the 
vicinity of Sgr A and the Arches; it is lower than average in dense molecular 
clouds that are bright at 500 \micron\ and are sometimes seen in extinction at 70 \micron.
Large scale variations in the 
spectral index are correlated with some features, but it's not clear that 
consistent trends are present. Figure \ref{fig:free_beta} shows a histogram 
of $\beta$ that reveals that most of the data are consistent with a mean 
emissivity index of $\beta \approx 2.25$. A portion of the tails in the histogram,
and some more extreme outliers, arise from scattered pixels where poor 
signal to noise in one or more bands leads to anomalous fits.
Figure \ref{fig:free_temp_beta} shows that we find no strong correlation 
of $T_{\rm d}$ and $\beta$ in this region.
We do note that regions with $T_{\rm d} \gtrsim 23$~K
generally have lower $\beta$ than most regions, although regions 
with relatively low $\beta$ can be found at all temperatures. 
The maps in the lower panels of Figure \ref{fig:free_temp_beta} show
that a relatively distinct clustering of points with 
high $T_{\rm d}$ and low $\beta$ are mostly found in the 
vicinity of Sgr A and the Arches. Regions with low $T_{\rm d}$ and 
low $\beta$ are mostly associated with molecular clouds such as the 
Sgr B2 complex.

\begin{figure*}[t] 
   \centering
   \vspace{-0.25in}
   \includegraphics[width=6.5in]{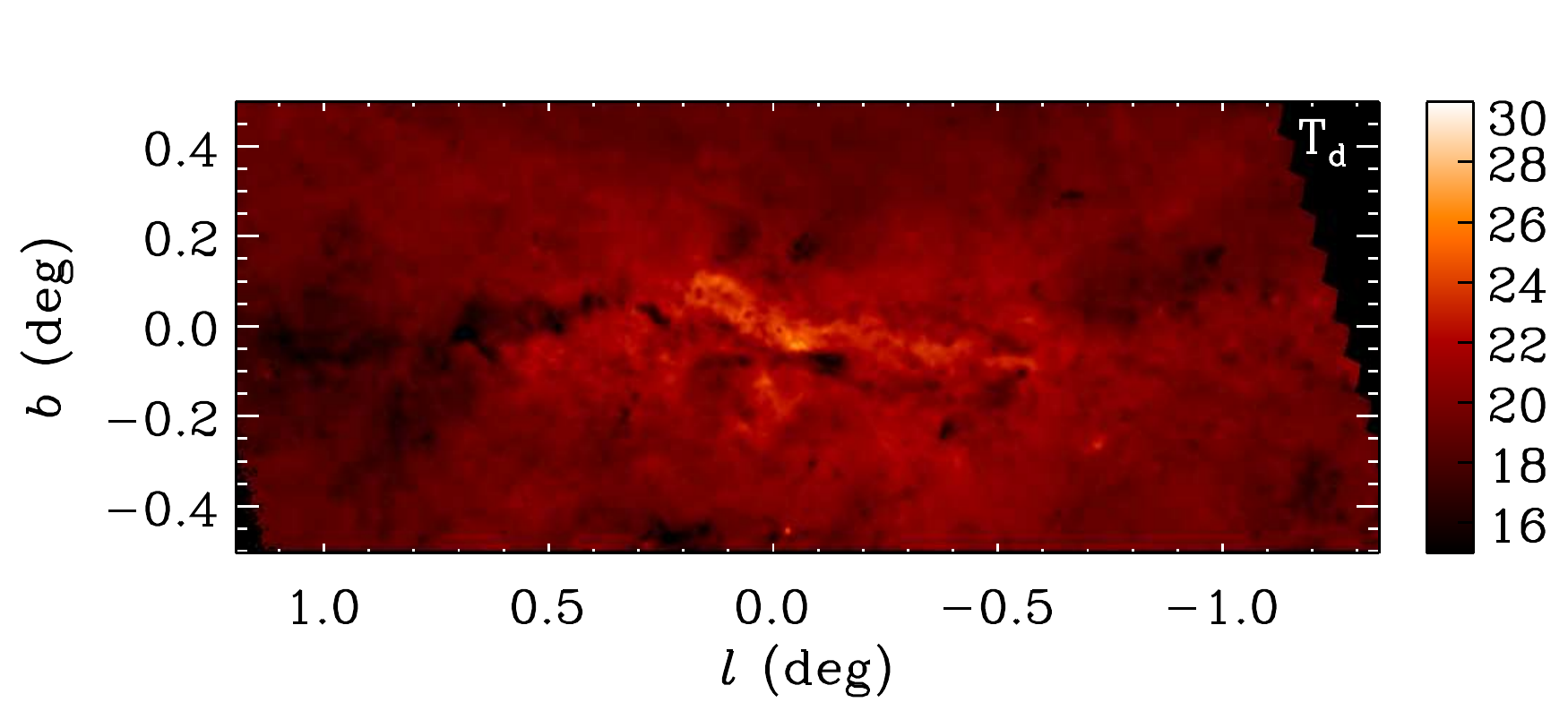}\\
   \includegraphics[width=6.5in]{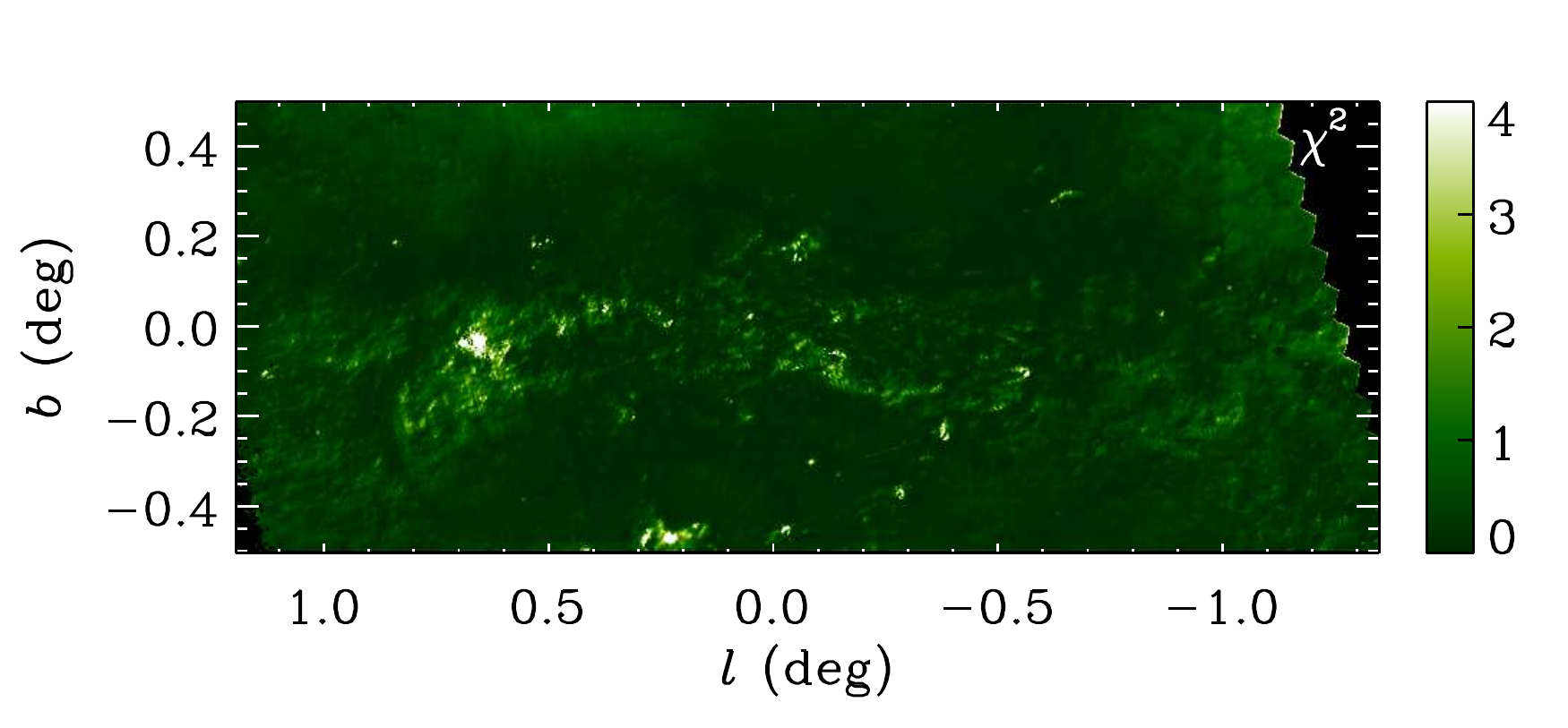}\\
   \includegraphics[width=6.5in]{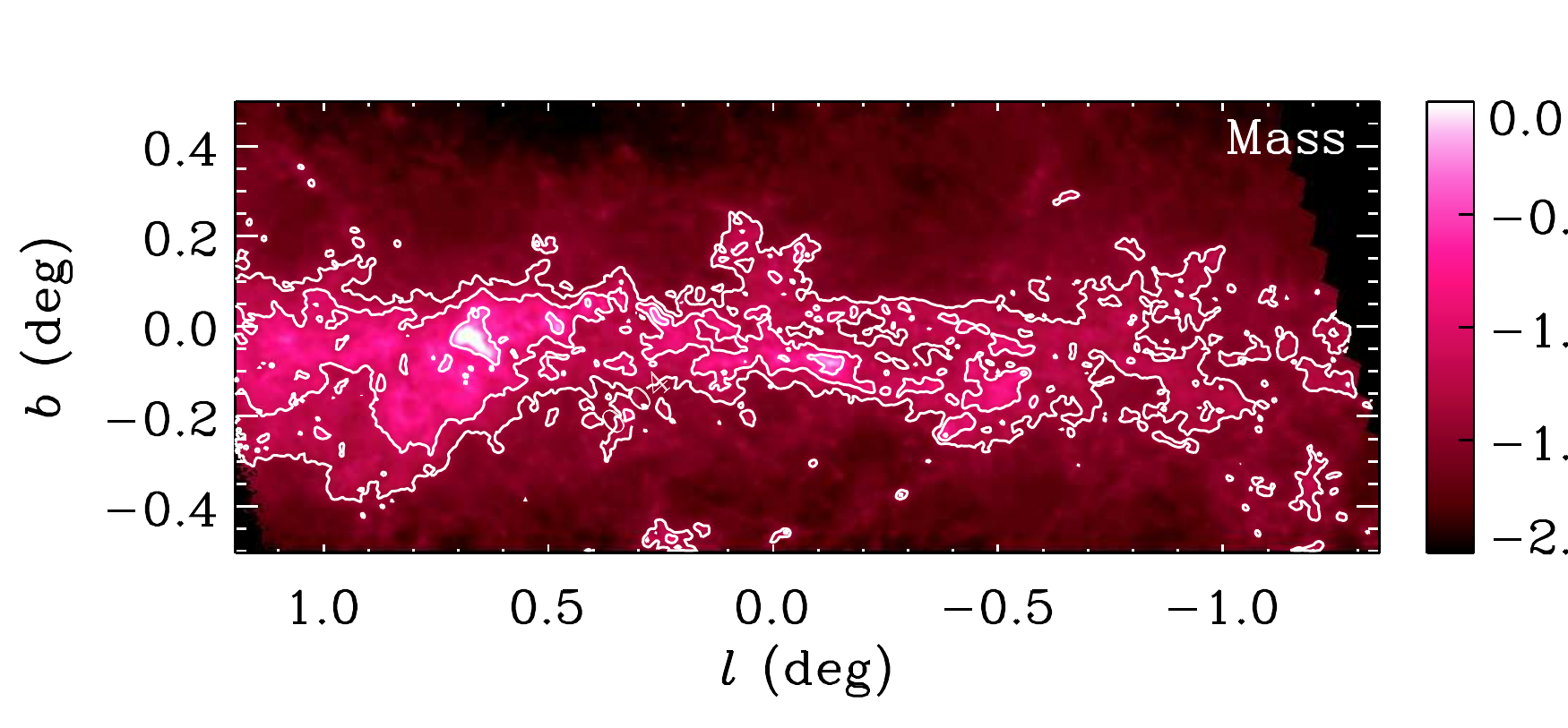} 
   \caption{Modeling the 160--500 $\mu$m images using a free dust temperature 
   $T_{\rm d}$ (top) but a fixed $\beta = 2.25$. The display range 
   is [15,35] for $T_{\rm d}$. The middle panel 
   shows $\chi^2$ for the fit at each pixel on a range of [0,4] 
   (fixed for comparison with other figures). Input data were convolved to 
   $37''$ resolution. $T_{\rm d}$ is very similar to the case where $\beta$ is 
   a free parameter, except now all the variations are smoothly continuous
   without any spurious pixel-to-pixel variations caused by noise. $\chi^2$
   becomes worse at some locations (cf. Figure \ref{fig:fit12}), 
   as expected for a more constrained model.  
   The bottom panel shows the model's dust mass surface density 
   (proportional to the normalization parameter $A$). The image is 
   logarithmically scaled from 0.01 -- 1~$M_{\sun}$~asec$^{-2}$.
   Contours are at [0.04, 0.08, 0.3]~$M_{\sun}$~asec$^{-2}$.
   \label{fig:fit14}}
\end{figure*}

Because of the occasionally spurious fitting results, we next 
investigated constrained 
fits where the emissivity index was fixed at $\beta = 2.25$. Figure \ref{fig:fit14} 
shows that the derived temperatures are similar to those found when $\beta$ is 
a free parameter, but now there are no spurious pixel-to-pixel variations. The 
values of $\chi^2$ increase somewhat, but there are few major changes.
This constrained fit is used as our standard model in further analysis.
The third panel in Figure \ref{fig:fit14} shows the dust mass surface 
density derived from the fit normalization $A$, assuming  
$\kappa_0 = 50$~cm$^{2}$~g$^{-1}$ at $\lambda_0 = 100$~$\mu$m 
and that all the dust is at the Galactic center distance of 8.18~kpc
\citep{Abuter:2019}.

\begin{figure*}[t] 
   \centering
   \includegraphics[width=6.5in]{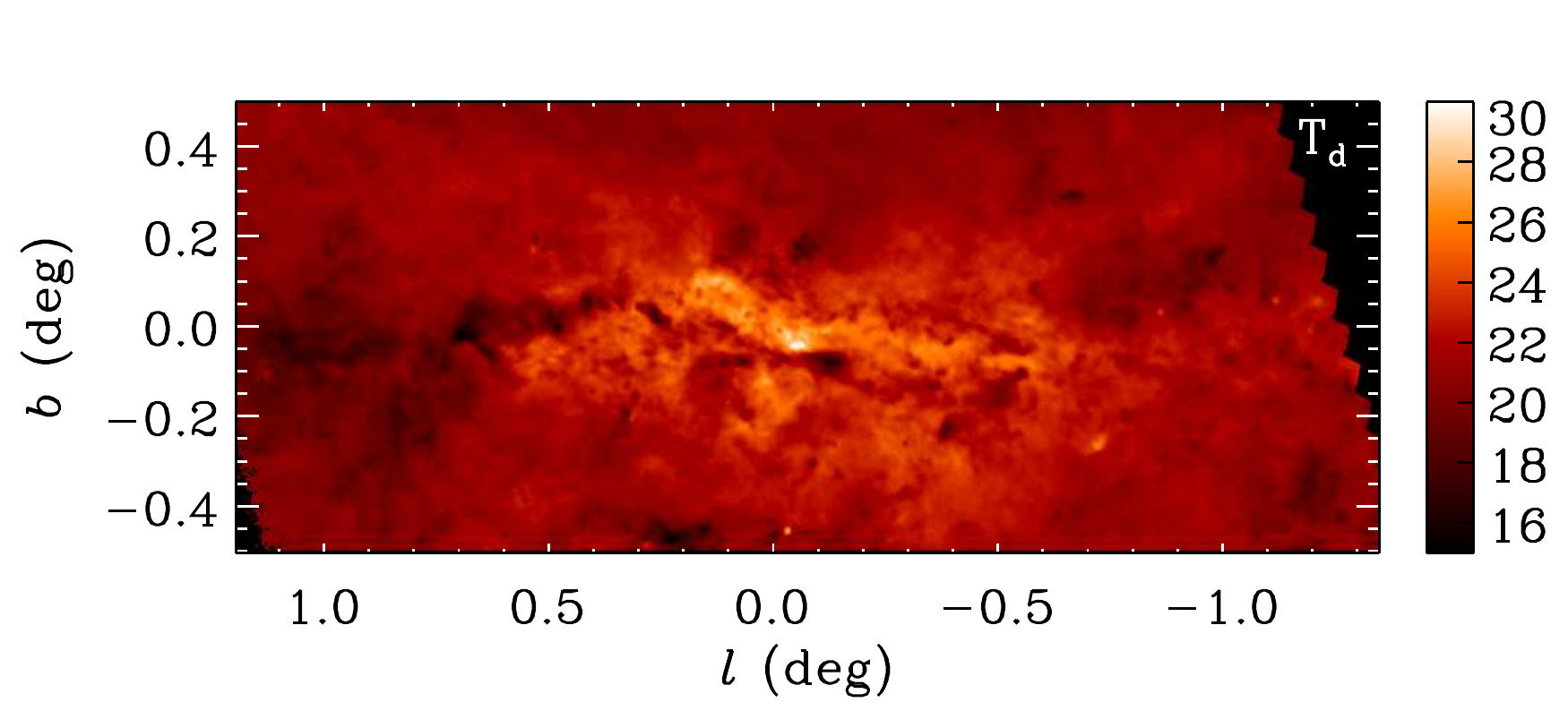}\\
   \includegraphics[width=6.5in]{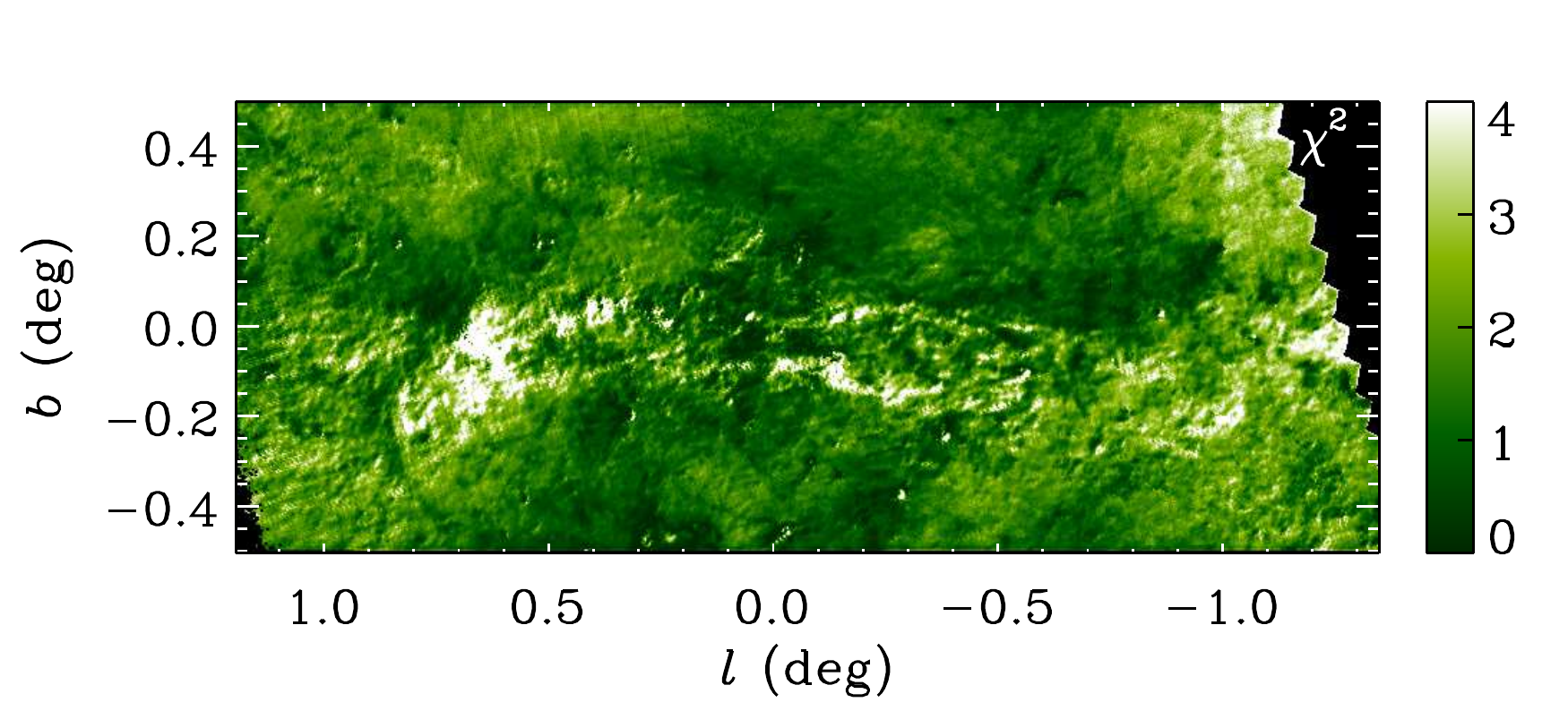} 
   \caption{Modeling the 160--500 $\mu$m images using a free dust temperature 
   $T_{\rm d}$ (top) but a fixed $\beta = 2.00$. The display range 
   is [15,35] for $T_{\rm d}$. The lower panel 
   shows $\chi^2$ for the fit at each pixel on a range of [0,4] 
   (fixed for comparison with other figures). Input data were convolved to 
   $37''$ resolution. $T_{\rm d}$ is similar to, but systematically higher 
   than the case where $\beta = 2.25$. $\chi^2$ is now much worse at all
   locations (cf. Figures \ref{fig:fit12} and \ref{fig:fit14}), indicating 
   that the {\it Herschel} data alone require $\beta > 2.0$. Curiously, it is 
   in this $\chi^2$ image that a circular feature at $(l,b) = (-0.35, 0.18)$ 
   is most easily distinguished.
   \label{fig:fit13}}
\end{figure*}

For comparison, we also performed a constrained fit using a 
more conventional value of $\beta = 2.0$. The temperature distribution
(Figure \ref{fig:fit13}) is very similar to that found for 
$\beta = 2.25$ except it is systematically 
slightly warmer, as would be expected for 
a flatter emissivity index. There are a few isolated regions where $\chi^2$
is distinctly lower for the flatter spectral index. Many of these are associated
with higher latitude features that appear as IR dark clouds at shorter 
wavelengths ($\lambda \lesssim 70$~$\mu$m).
The high latitude and extreme darkness of these clouds suggest that they are in 
the foreground relative to the Galactic center. 

\begin{figure*}[t] 
   \centering
   \includegraphics[width=2.25in]{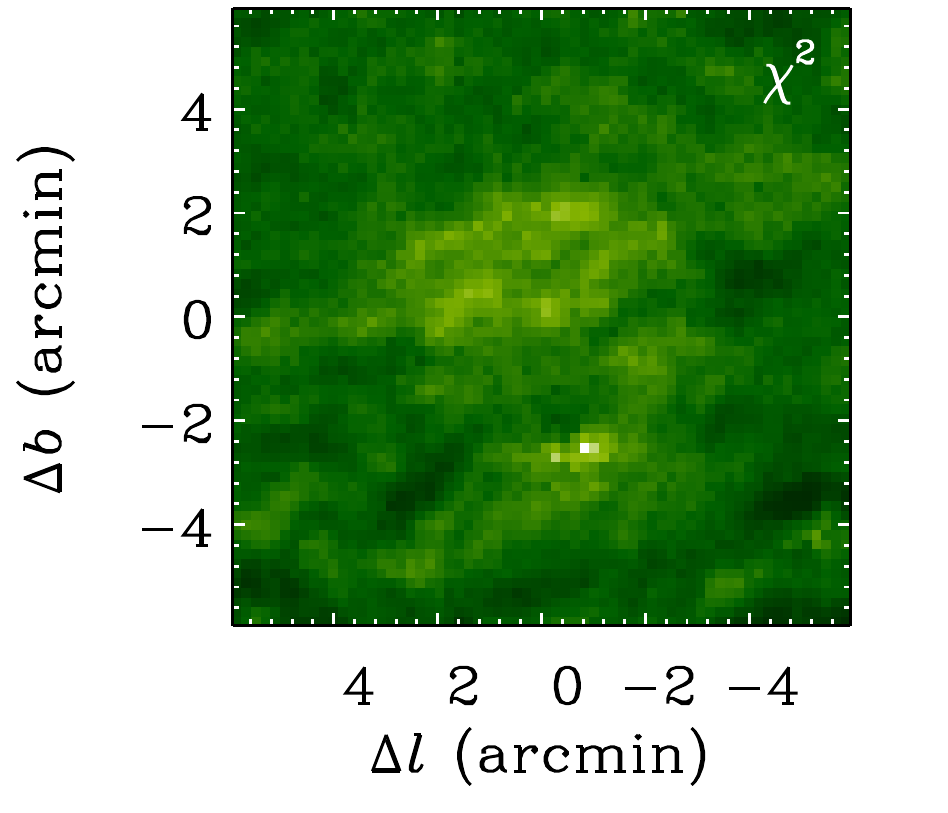}
   \includegraphics[width=2.25in]{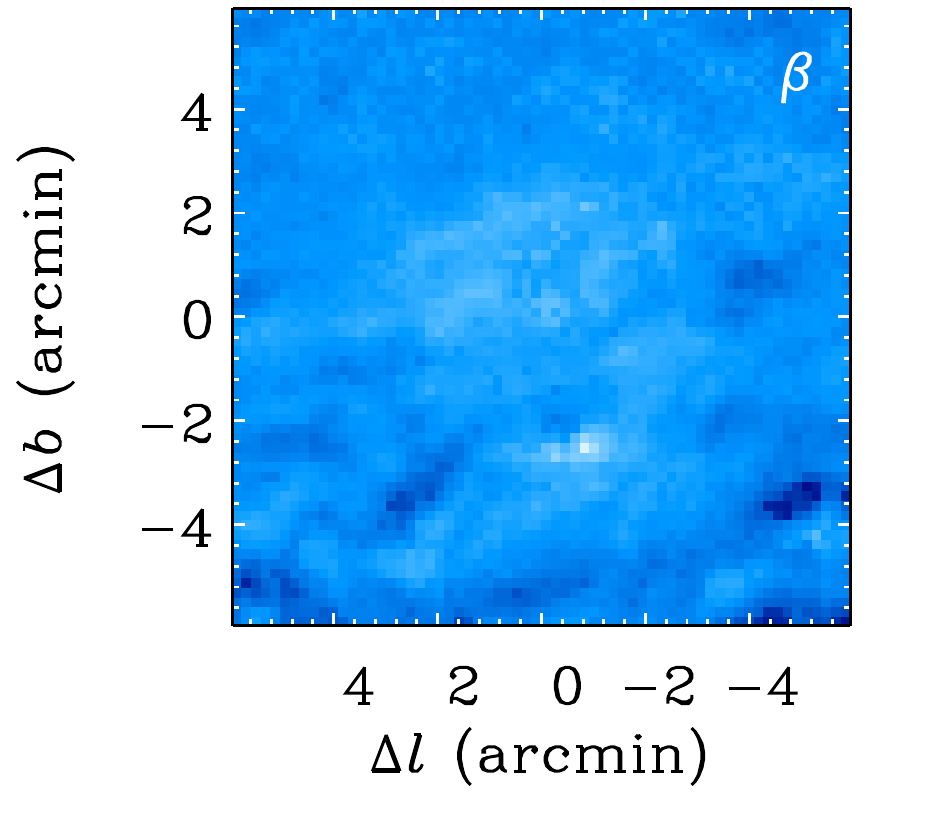} 
   \includegraphics[width=2.25in]{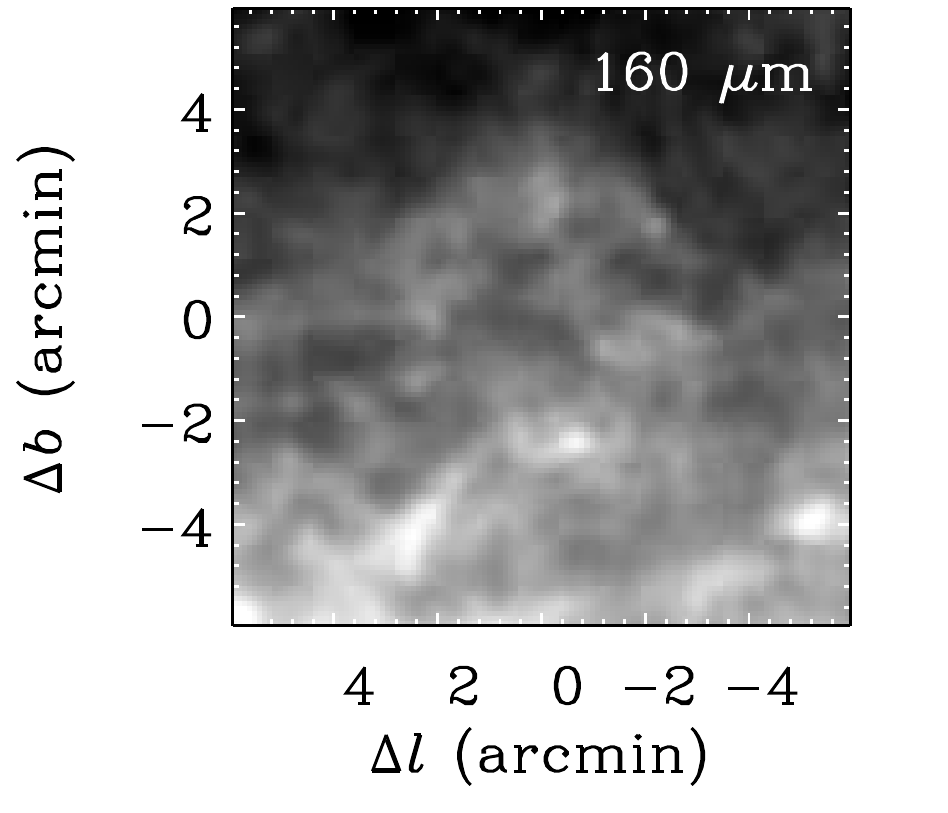} 
   \caption{Magnified images of an unusual circular structure 
   as revealed by fitting deviations ($\chi^2$), emissivity index ($\beta$),
   and 160~$\mu$m emission (left to right). The structure is centered at $(l,b) =
   (359.65,+0.18)$. The structure seems to be largely defined by unusual dust 
   properties (See Section \ref{sec:far-ir}).
   \label{fig:circle}}
\end{figure*}

There is a peculiar shell-like 
feature in the $\chi^2$ map for $\beta = 2.0$ (Figure \ref{fig:fit13}).
It is $3'$ in radius at $(l,b) = (359.65,+0.18)$ and
is also distinguished from its surroundings by a slightly steeper 
value of $\beta \sim 2.35$ (Figure \ref{fig:fit12}).
Figure \ref{fig:circle provides a magnified view of this region.}
The structure is not particularly distinctive, but can be discerned, 
in the {\it Herschel} 160 - 500 $\mu$m intensity maps. 
The structure is not evident in {\it Spitzer} images at 3.6 - 24 $\mu$m
or at radio wavelengths. While there are many possible candidates 
for a central star to this structure, none are particularly distinguished
by exceptional colors or brightnesses at 1.25 - 24 $\mu$m. 

\subsection{2~mm and 70 \micron\ emission}		

\begin{figure*}[t] 
   \centering
   \includegraphics[width=6.5in]{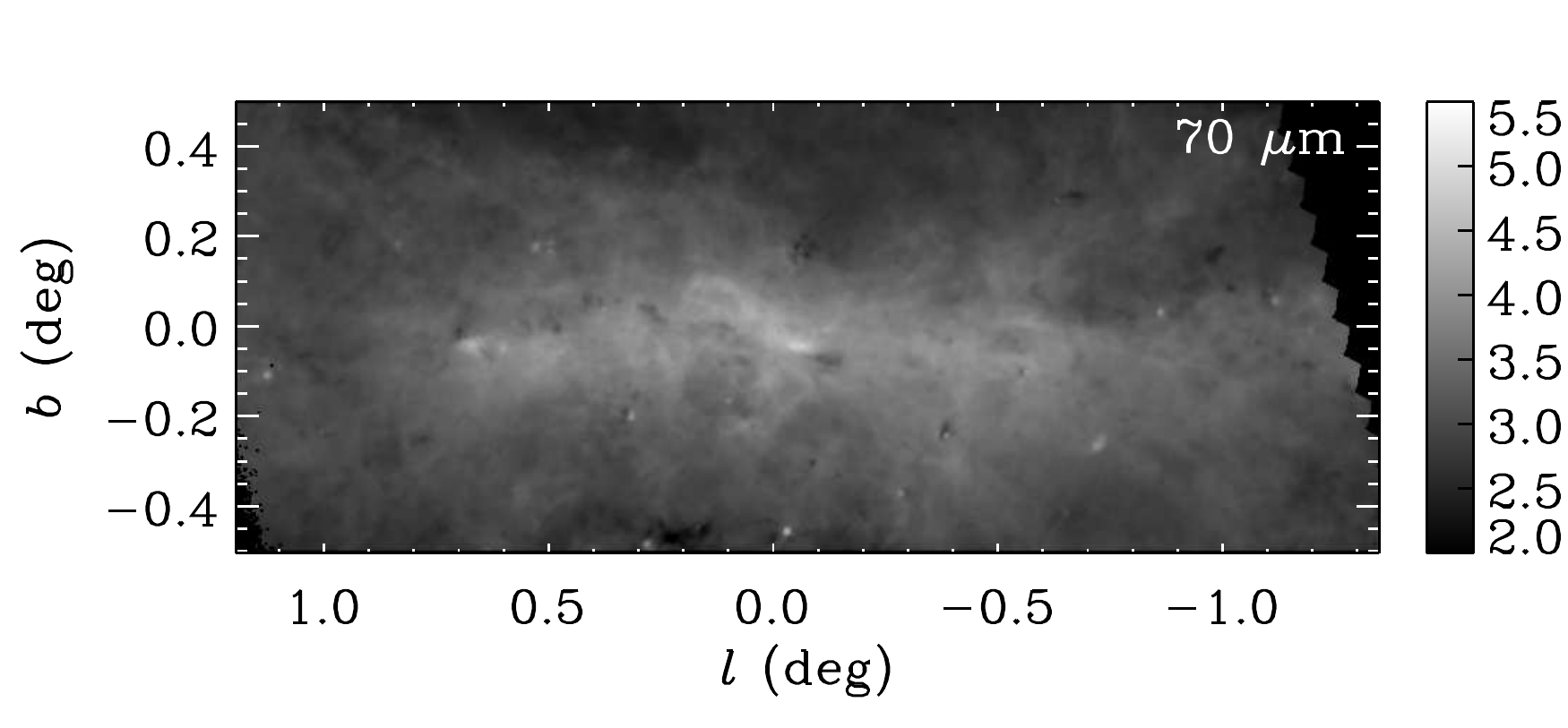}\\
   \includegraphics[width=6.5in]{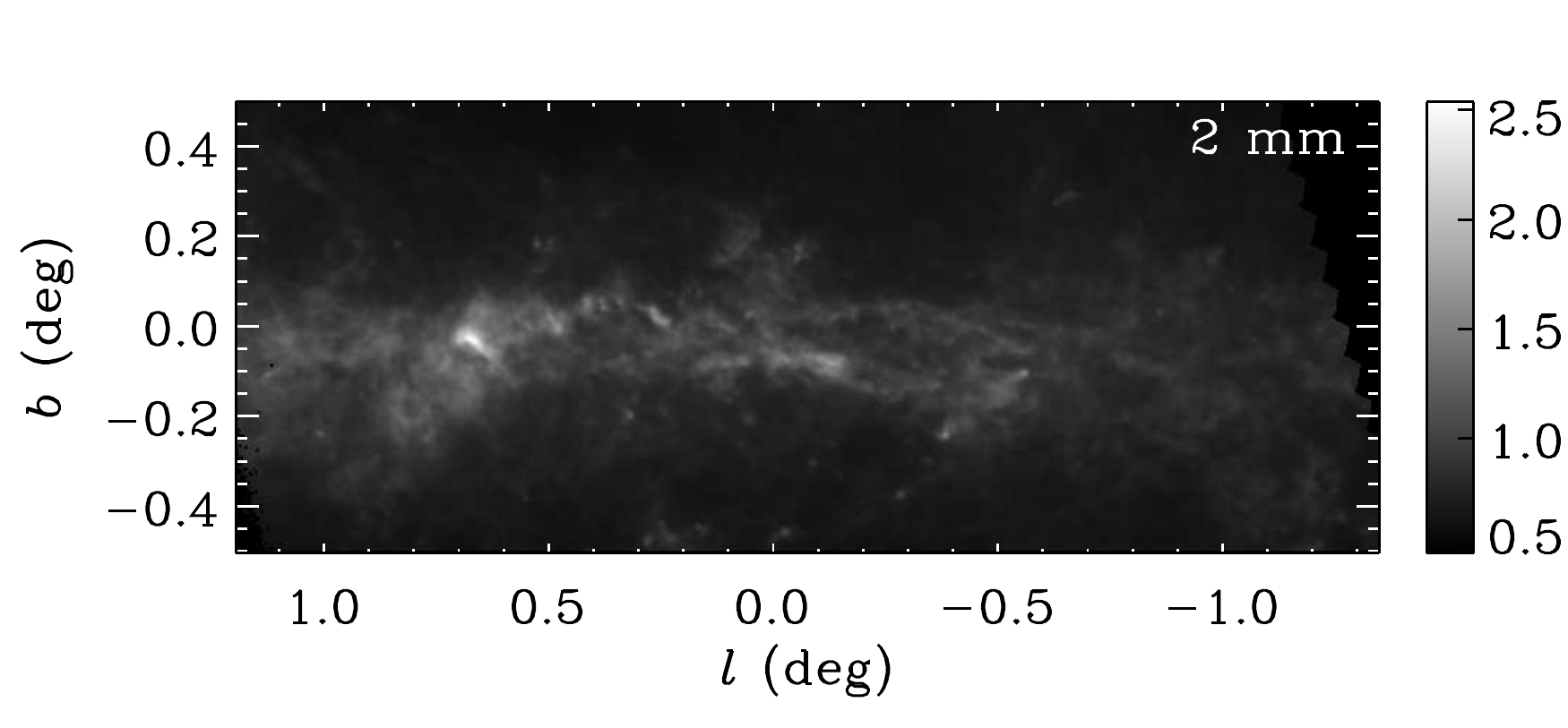} 
   \caption{Images of 70 $\mu$m (top) and 2~mm (bottom) emission extrapolated 
   from the fixed $\beta = 2.25$ model of the 160--500 $\mu$m emission.
   The data are convolved to $37''$ resolution. 
   These images are visually similar to the observed intensities shown 
   in Figures \ref{fig:pacs} and \ref{fig:gismo}, and are shown on the 
   same logarithmic display ranges.
   \label{fig:extrap14}}
\end{figure*}

We extrapolate the emission model fit to the 160 -- 500 \micron\ data to 
predict the dust emission at 2~mm and at 70 \micron. The extrapolated intensity 
maps are shown in Figure \ref{fig:extrap14}. 

\begin{figure*}[t] 
   \centering
   \includegraphics[width=6.5in]{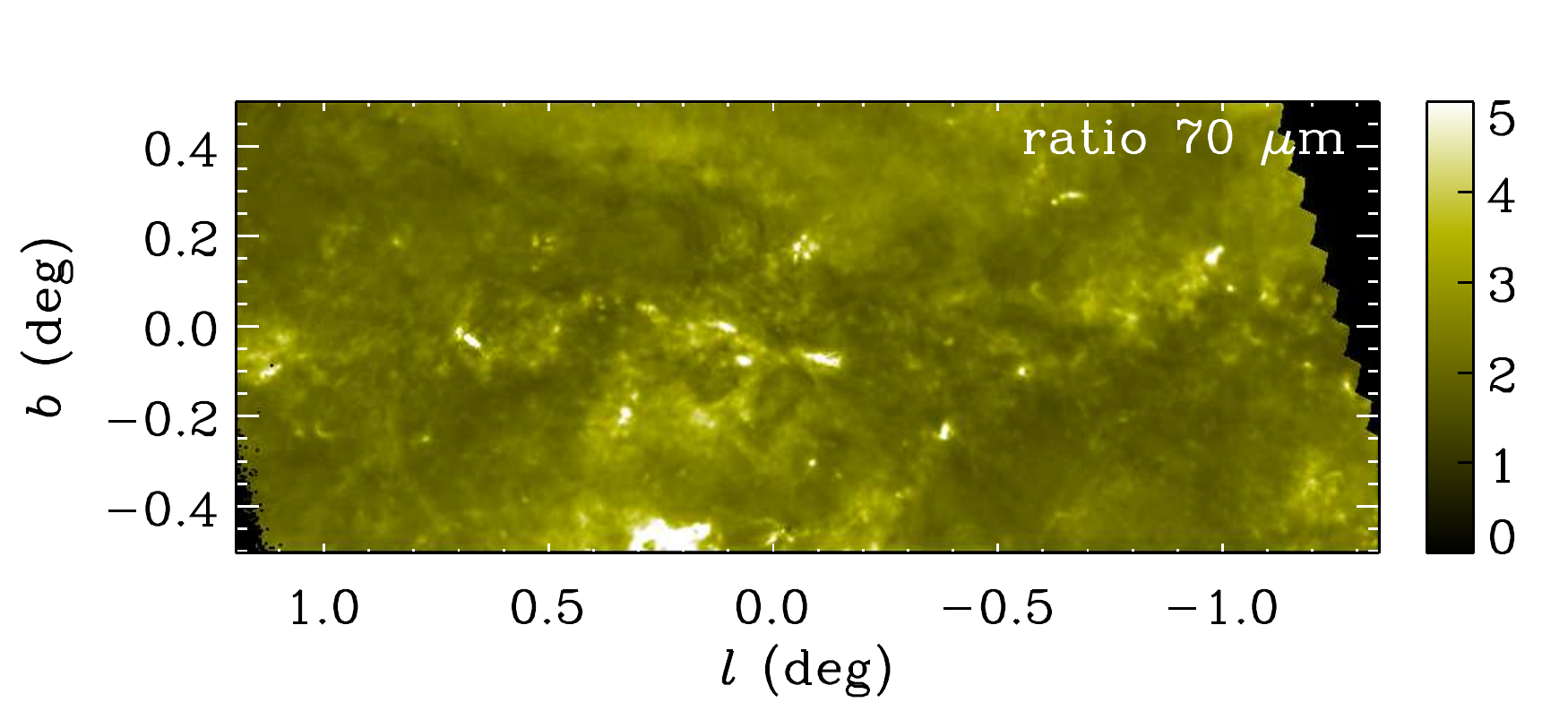}
   \caption{Ratio of observed to extrapolated 70 \micron\ intensities
   (from Figures \ref{fig:pacs} and \ref{fig:extrap14}). 
   The linear grayscale range is [0,5]. White regions, 
   where the observed 70 $\mu$m intensity is significantly greater 
   than the extrapolated intensity, include star-forming regions with 
   warm dust and, prominently, cold molecular clouds and IRDCs. 
   \label{fig:ratio70}}
\end{figure*}

The observed 70 \micron\ emission is always larger than the 
70 \micron\ intensity extrapolated from the model (Figure \ref{fig:ratio70}). The median 
ratio of the observed to predicted intensities is 2.1. The ratio
tends to be higher in regions where the temperature is low,
even though such regions can be optically thick at 70~$\mu$m and
appear as IR dark clouds. 
The underestimate is likely due to two factors. The first is the 
neglect of a distribution of dust temperatures in the model. 
Warmer dust (either smaller dust and/or stochastically heated dust)
will add intensity at the shorter wavelengths, broadening the spectrum. 
The second factor is the adoption of $\beta = 2.25$, which leads to a 
more sharply peaked spectrum than flatter values of the emissivity index. 
While the steep index is found to fit the longer wavelengths, it is not
necessary that this index continues to apply at shorter wavelengths. 
In either case, a more complete and physical model of the ISM emission
would require additional parameters to fit the 70 \micron\ data. However
at 160 -- 500 \micron\ wavelengths, the fit is already sufficiently good 
that additional parameters are not statistically warranted.

\begin{figure*}[t] 
   \centering
   \includegraphics[width=6.5in]{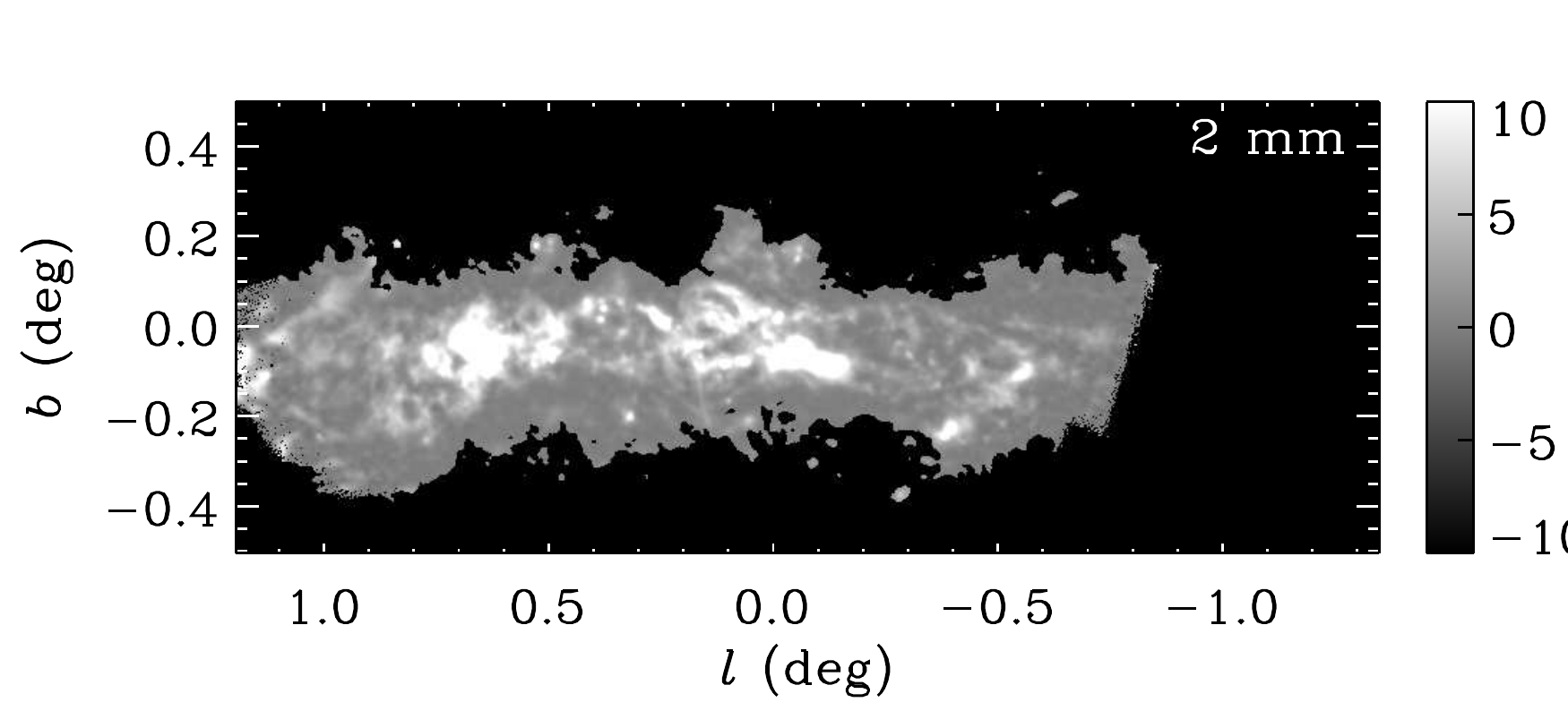}\\
   \includegraphics[width=6.5in]{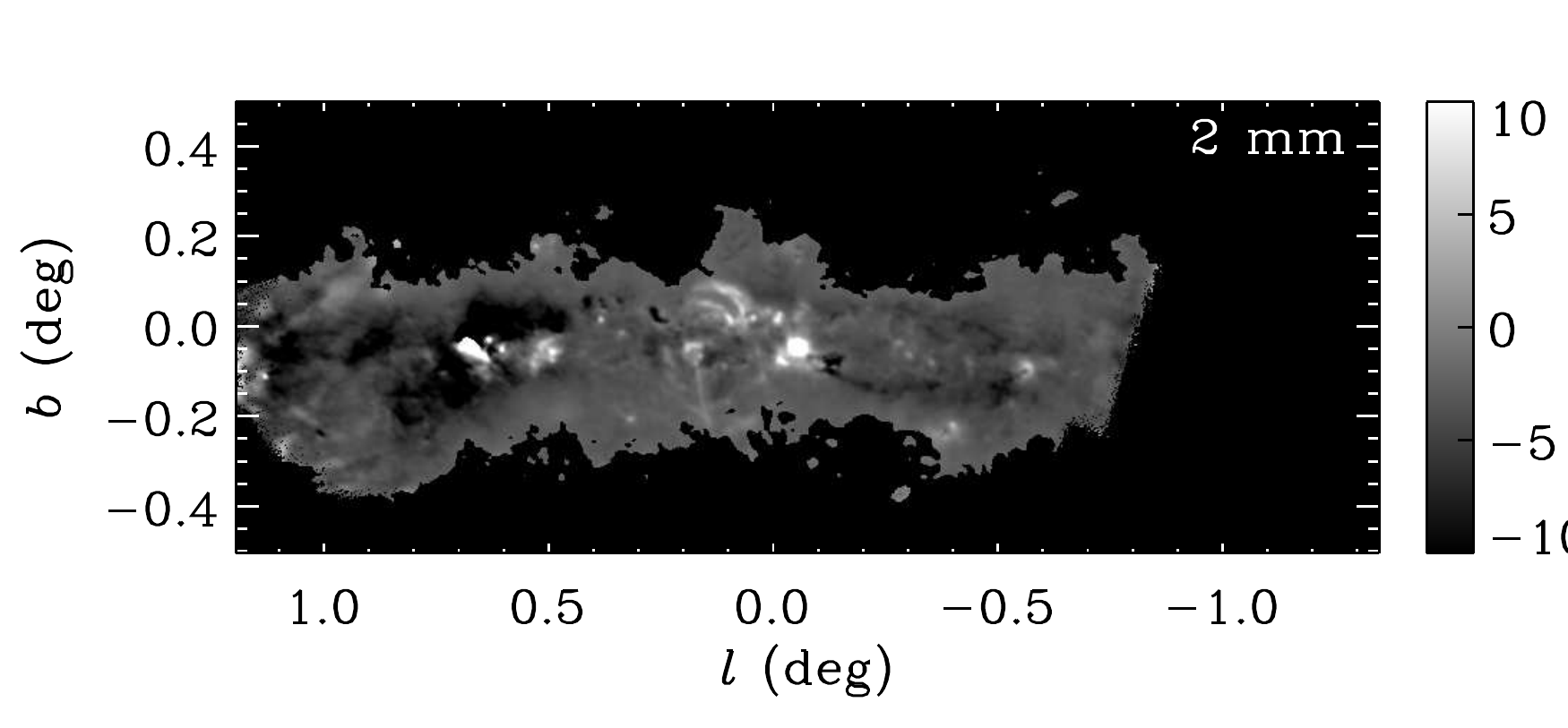} 
   \caption{The observed 2~mm intensity (top) displayed on a linear
   grayscale range of [-10,10] MJy sr$^{-1}$, 
   and the difference (bottom) between the observed and 
   extrapolated 2~mm intensities on the same display range ($37''$ resolution).
   The difference image shows that most of the 2~mm emission is associated 
   with thermal emission from dust. In regions where $T_{\rm d}$ is low, the 
   emission is sometimes over-subtracted (appearing dark). Regions exhibiting 
   2~mm emission in excess of the dust 
   emission (bright in the difference image) are \ion{H}{2} regions and ionized structures that radiate via 
   free-free emission at radio wavelengths, and non-thermal emission from 
   the central filament of the Galactic center Radio Arc.
   \label{fig:excess2mm}}
\end{figure*}

Despite the steep emissivity index, the model generally overestimates the 
2~mm emission, especially in the cold dense molecular clouds of Sgr B2 and the 
Galactic center (Figure \ref{fig:excess2mm}). However, there are many structures where 
the observed 2~mm emission is in excess of the extrapolated dust emission.
These include: extended and compact regions in Sgr B2, and Sgr B1; 
the Arches, Sickle, and Pistol Nebula; The Sgr A region; and other 
regions near Sgr C. Additionally, the brightest of the nonthermal 
filaments in the Galactic center is also evident, especially to the south 
of the Sickle nebula. There is very weak evidence of emission from some 
of the other adjacent parallel filaments.
However, none of the other nonthermal filaments are 
detected at other locations around the Galactic center.

\section{Discussion} \label{sec:discussion}

The far-IR spectral index of $\beta = 2.25$ derived here is
steeper than would be expected for typical carbon grains ($\beta=1$) 
or silicate grains ($\beta = 2$). However the relatively high value 
of $\beta$ is similar to that derived in other studies that fit
for dust properties in selected areas or over the whole sky.
For example, \cite{Planck-Collaboration:2014} used {\it Planck} 
observations at $30'$ resolution to fit 
dust emission over the whole sky. They found $\beta\approx 1.6$ 
at high Galactic latitudes ($|b|\gtrsim 10\arcdeg$), but increasing
to $\beta = 1.8 - 2.0$ in the inner Galactic plane ($|l|\lesssim 30$).
Using 450 and 850~$\mu$m SCUBA observations of the CMZ 
combined with 100~$\mu$m IRAS data, \cite{Pierce-Price:2000} 
found vales of $\beta \sim 2.4$ when $\beta$ was allowed to be a free
parameter of their model. 
\cite{Paradis:2010} used {\it Herschel} data at 160, 250, 350, 
and 500 $\mu$m, and {\it IRAS} (IRIS) 100 $\mu$m data 
\citep{Miville-Deschenes:2005} to model dust in 
two fields at $l = 30\arcdeg$ and $l = 59\arcdeg$.
Their derived results at $l = 30\arcdeg$ have a similar mean 
$T_{\rm d}$ and $\beta$ as ours, but show a much stronger 
inverse correlation between these two parameters.
The lack of correlation towards the Galactic center may be caused 
by more intrinsic variation in dust properties along longer 
lines of sight. The lack of spatial smoothness 
in the derived $T_{\rm d}$ and $\beta$ (Fig.~\ref{fig:fit12}), and the sometimes spurious 
results, may be the response to noise in the data 
when using $\chi^2$ fitting methods \citep{Juvela:2012,Juvela:2013}.

The total dust masses within the 0.04, 0.08, and 0.3 ~$M_{\sun}$ asec$^{-2}$
contours shown in Figure \ref{fig:fit14} are
4.3, 3.5, and $0.99\times10^5$~$M_{\sun}$
when 0.04~$M_{\sun}$ asec$^{-2}$ is subtracted as background. 
With no background subtraction, the dust masses are 
7.8, 4.6, and $1.05\times10^5$~$M_{\sun}$, and the total dust mass in the 
shown image is $1.3\times10^6$~$M_{\sun}$. 
Integrating the dust mass within the 0.08 $M_{\sun}$ asec$^{-2}$ contour 
and where $0.75>l>-0.65$ yields $2.4\times10^5$~$M_{\sun}$
over an area of $4.2\times10^{-5}$~sr [consistent with the area
and the $3\times10^7$~$M_{\sun}$ gas mass reported for the 
twisted ring structure by \cite{Molinari:2011}].

It was expected that extrapolation of the single-temperature dust model to 70 $\mu$m would 
underestimate the emission in star-forming regions where warm dust is common.
This is indeed found in our modeling. 
However, underestimates also are unexpectedly found at the 
locations of IRDCs and other low temperature regions 
(Figure \ref{fig:ratio70}). This result indicates 
that the 70 $\mu$m emission is coming from a substantially warmer dust component,
possibly smaller or stochastically heated grains, or from warmer dust along the 
line of sight. 

The extrapolation of the model to 2~mm shows that the observed emission can be 
brighter or fainter than the model prediction (Figure \ref{fig:excess2mm}).
The model tends to over-predict emission in the giant molecular clouds of 
Sgr B2, and the twisted ring of molecular clouds \citep{Molinari:2011}. These
are also well-correlated with regions of low dust temperature, suggesting 
a value of $\beta>2.25$ would be applicable to the colder clouds. This is 
weakly evident in the initial model which allowed a freely varying 
$\beta$, but is obscured by the overall noisiness of that unconstrained model
(Figs. \ref{fig:fit12}, \ref{fig:free_temp_beta}). An additional contribution to overestimates of 
the 2~mm emission may be caused by missing large-scale structure in the GISMO
image (\S\ref{ssec:gismo}), although the small scale of some overestimated features indicates that 
this cannot be the sole explanation. 

Finally, the real regions of interest are where there is 2~mm 
emission in excess of the model prediction. These regions are places where
the 2~mm observations are sensitive to emission components other than the 
thermal emission from dust. For the most part, the excess regions correspond
to the known star-forming regions and ionized structures in the Galactic
center. The most prominent are Sgr A and its nearby \ion{H}{2} regions Sgr A A-D 
(these 4 compact \ion{H}{2} regions are not fully resolved from one another),
the Arches, the Sickle and the Pistol Nebula, Sgr B1, and Sgr C. A large and 
bright region of excess emission is located around and to the south of the 
compact \ion{H}{2} regions embedded in Sgr B2. Most surprisingly, there is
excess 2~mm emission from the central filament of the Radio Arc. The emission
extends southward from the vicinity of the Sickle and Pistol Star, 
$(l,b)=(0.16,-0.06)$, to a location near $(l,b)=(0.14,-0.21)$. 
The Radio Arc is the largest and brightest of the collection of non-thermal 
filaments that have been identified in the Galactic center. 
A detailed investigation of this feature and the 2~mm emission from 
other non-dust sources is presented by \citet[][Paper II]{Staguhn:2019}.

In \ion{H}{2} regions, the dust emissivity, 
$\epsilon_\nu^{\rm{IR}}(\lambda)$, is given by  
\begin{equation}
\epsilon_\nu^{\rm{IR}}(\lambda) = 4 n_{\rm d}\ m_{\rm d}\ \kappa(\lambda)\ \pi B_\nu(T_{\rm d},\lambda)
\end{equation}
where $n_{\rm d}$ is the number density of dust grains, 
$m_{\rm d}$ is the mass of a single dust grain, $\kappa(\lambda)$ is the mass absorption 
coefficient for the dust, and $B_\nu(T_{\rm d},\lambda)$ is the Planck function 
evaluated at the dust temperature, $T_{\rm d}$.
Meanwhile, the free-free emissivity, $\epsilon_\nu^{\rm{ff}}(\lambda)$, is given by:
\begin{equation}
\epsilon_\nu^{\rm{ff}}(\lambda) = C\ g_{\rm ff}\ \frac{n_{\rm e}\ n_{\rm i}}{T^{0.5}\ e^{h\nu/kT}}
\end{equation}
where $C$ is a numerical constant ($\approx 5.44\times10^{-41}$), 
$g_{\rm ff}$ is the gaunt factor, $n_{\rm e}$ and $n_{\rm i}$ are the
number densities of electrons and ions, and $T$ is the gas temperature.
Thus for a dusty region of ionized gas, the observed ratio, $R$, of 70~$\mu$m emission 
from $T_{\rm d}=51$~K dust \citep{Kaneda:2012} 
to 2~mm free-free emission from $T=10^4$~K gas can 
be used a rough diagnostic of the electron density via 
\begin{equation}
R \approx 1.7\times10^6\ n_{\rm e}^{-1}
\end{equation}
assuming a dust-to-gas mass ratio of $Z = 0.01$.
If the gas temperature were instead taken to be 8000~K, then the electron 
density derived from the ratio of IR to free-free emission 
would increase by $\sim10\%$.
The coefficient in Equation 9 is much more sensitive to $T_d$,
dropping by a factor of $\sim70$ if $T_d = 25$~K.

\begin{figure*}[t] 
   \centering
   \includegraphics[width=6in]{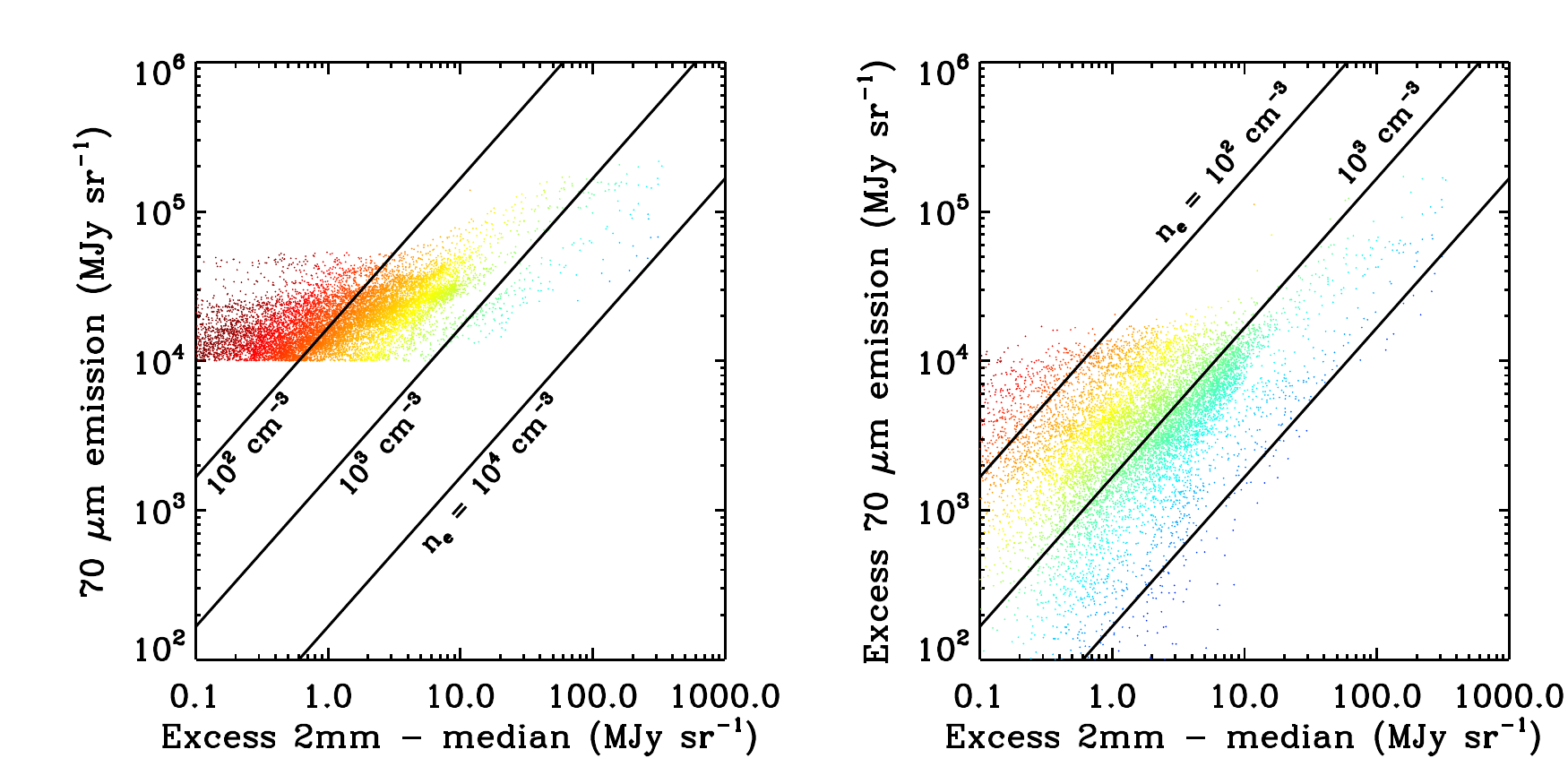}
   \caption{Ratio of residual 2~mm emission (mostly free-free) to 
   70~$\mu$m emission without (left) and with (right) subtraction 
   of the emission of cold dust extrapolated from longer wavelengths.
   The points are shaded as a function of $n_{\rm e}$, with several lines of 
   constant density overplotted for guidance.
   \label{fig:2-70}}
\end{figure*}

\begin{figure*}[t] 
   \centering
   \includegraphics[width=6.5in]{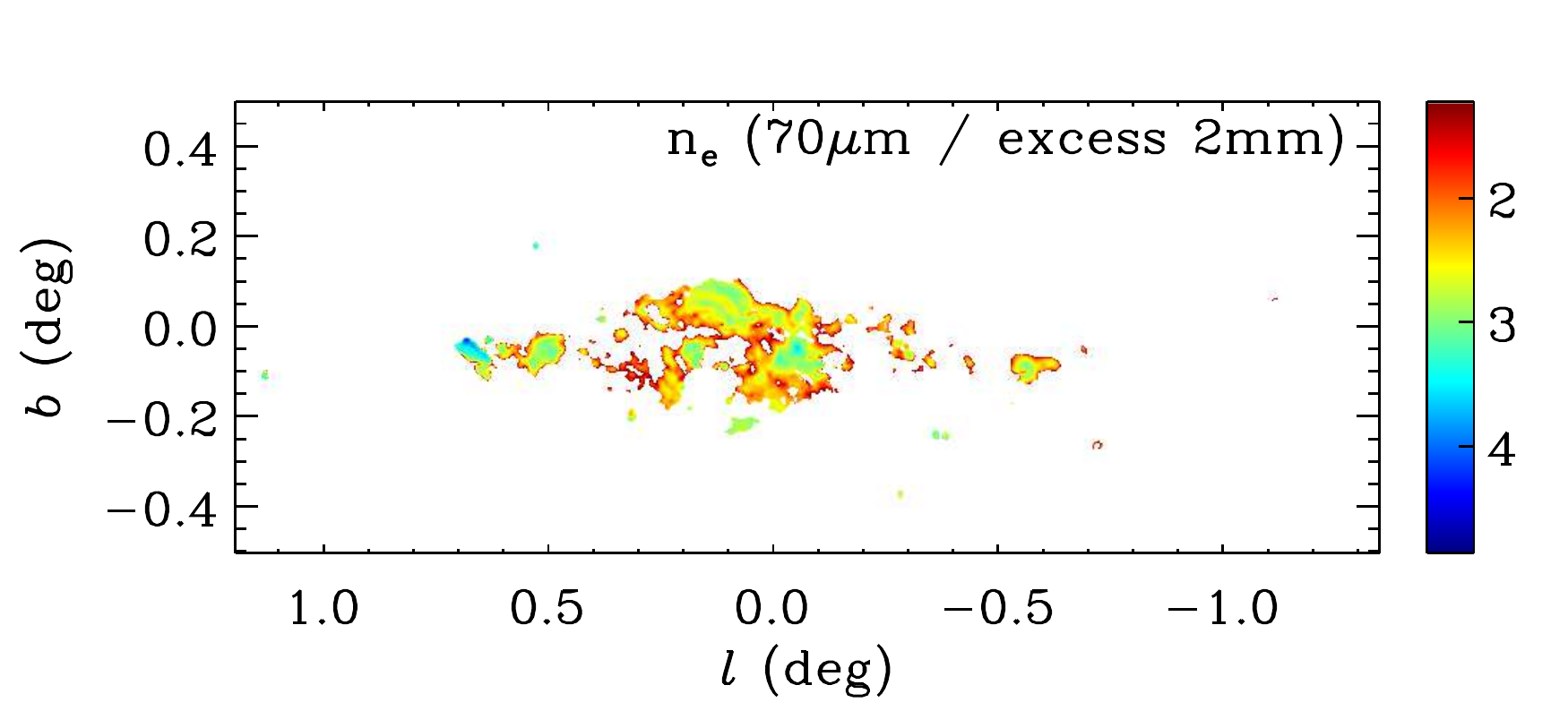}\\
   \includegraphics[width=6.5in]{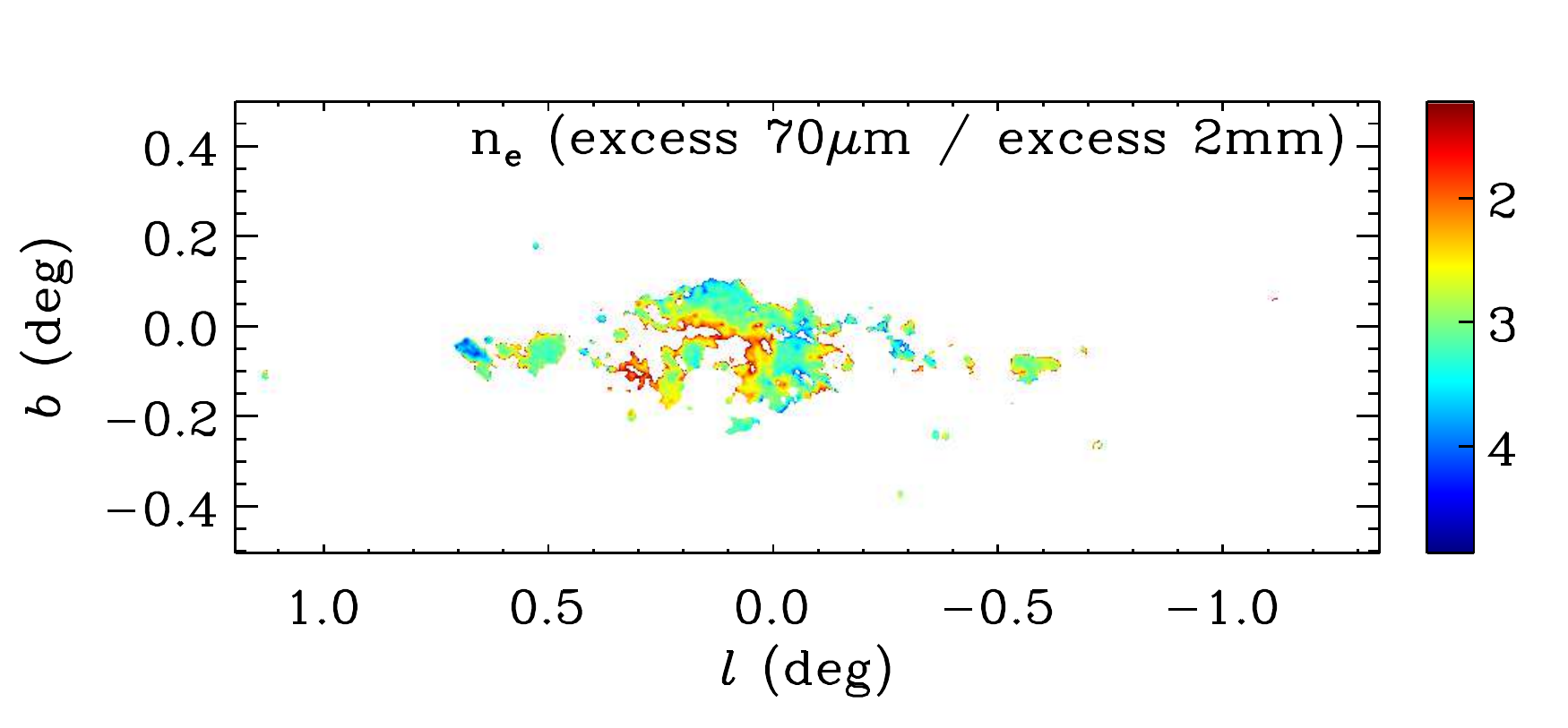}
   \caption{Maps of the logarithm of the electron density, $\log(n_{\rm e})$, 
   implied by the ratio of residual 2~mm emission (mostly free-free) to 
   70~$\mu$m emission without (top) and with (bottom) subtraction 
   of the emission of cold dust extrapolated from longer wavelengths.
   The latter case better isolates dust emission from warm regions where
   free-free emission is more likely to arise, and results in derived $n_{\rm e}$ 
   values that are less correlated with the 70 $\mu$m brightness.
   The color scale for $\log(n_{\rm e})$ is the same as in    
   Figure~\ref{fig:2-70}.
   \label{fig:ff_dust}}
\end{figure*}

To employ this relation as a density diagnostic, we have plotted the
2~mm free-free emission (i.e. the residual 
after subtraction of the ISM dust contribution), against 
the 70~$\mu$m emission in Figure \ref{fig:2-70}.
We compare with the observed 70~$\mu$m emission, and with 
the emission after the subtraction of the extrapolated emission 
of cold dust component determined from the 160-500~$\mu$m 
observations. The subtraction of this colder dust component is intended
to better isolate the warmer emission from \ion{H}{2} regions 
which produce the free-free emission.
The comparison is limited to regions where 
$I_\nu(70\mu{\rm m}) > 10^4$~MJy~sr$^{-1}$, and where the 
residual $I_\nu(2{\rm mm}) > 0$~MJy~sr$^{-1}$ after adjustment for the 
(negative) median background level. These scatter plots are 
shaded as a function of $n_{\rm e}$ as indicated by the diagonal lines.
Figure \ref{fig:ff_dust} maps out the locations of the points in 
Figure \ref{fig:2-70} using the same color coding for the density $n_{\rm e}$.
Without subtraction of the cold dust contribution at 70~$\mu$m, 
the relatively small variation in the selected 70~$\mu$m data means that 
the implied density is largely determined by the residual 2~mm brightness.
For example the extremely bright regions in Sgr B2 imply the highest densities. 
The less bright regions at Sgr B1, the Arches, and Sgr A 
have somewhat lower implied densities.
If the extrapolated emission of the colder dust is subtracted from the 
70~$\mu$m emission, the implied densities near Sgr B2 and Sgr A remain
high, with a slight increase. Implied densities are more 
strongly increased for the fainter regions, which still 
exhibit large-scale structure though the structure is no longer 
directly correlated with the observed intensity.

Electron densities are more typically estimated from the ratios
of ground state fine structure lines. Such determinations may be affected by 
extinction effects, but are only very weakly dependent on the gas temperature.
\cite{Rodriguez-Fernandez:2005} derived $n_{\rm e} <30 -200$~cm$^{-3}$
from the [\ion{O}{3}] 25/88 $\mu$m line ratios observed with {\it ISO}
in several lines of sight in the CMZ. \cite{Simpson:2018a} used the FIFI-LS
instrument on SOFIA to observe this line ratio in the Sgr B1 region, 
finding $n_{\rm e}\sim300$~cm$^{-3}$, with some denser regions along 
some edges of Sgr B1.
Using {\it Spitzer} observations of [\ion{S}{3}] 18.7/33.5 $\mu$m line ratios, 
\cite{Simpson:2018} typically found $n_{\rm e}\sim300$~cm$^{-3}$ in large 
regions across the CMZ, and $n_{\rm e}\sim300-600$~cm$^{-3}$ across the Arches 
region in particular \citep{Simpson:2007}.
Direct comparison with the Arches results shows our densities
are higher by factors of 1-10. Better agreement could be achieved by 
adjusting the estimates of one or more of the parameters in Equations (7) or (8),
but without further information it is impossible to judge which of the 
parameters should be adjusted.

\section{Summary} \label{sec:summary}

GISMO 2~mm observations of the Galactic center are dominated by the 
thermal emission of dust in the general ISM and particularly in 
molecular clouds. We model the far-IR emission seen by {\it Herschel} 
in order to predict and remove dust emission from the 2~mm image.
The dust temperature varies about a mean value of $T_{\rm d} \approx 19$~K.
Even though we find and set a relatively steep spectral index for 
the dust emission, $\beta = 2.25$, the 2~mm emission extrapolated 
from shorter-wavelength measurements is often overestimated for 
cold molecular clouds. This is 
consistent with previous studies finding an inverse correlation between
spectral index and dust temperature. Star-forming regions and other ionized
structures show additional 2~mm emission arising from the free-free mechanism.
Nonthermal emission from the central filament of the Galactic center 
Radio Arc is also detected at
2~mm. This is the shortest wavelength at which this feature has been detected.

\acknowledgments
{We would like to thank 
Carsten Kramer, Santiago Navarro, David John, Albrecht Sievers,
and the entire IRAM Granada staff for their support during the instrument 
installation and observations. 
We thank the referee for comments that improved the clarity and 
utility of the manuscript.
IRAM is supported by INSU/CNRS (France), 
MPG (Germany), and IGN (Spain). This work was supported 
through NSF ATI grants 1020981 and 1106284.}

\facilities{{\it Herschel} (PACS and SPIRE), IRAM:30~m (GISMO), VLA} 

\software{CRUSH \citep{Kovacs:2008}, IDLASTRO \citep{Landsman:1995}}

\bibliography{gismo_gc}

\end{document}